\RequirePackage{fix-cm}

\documentclass[twocolumn]{svjour3}       
\smartqed  
\usepackage{graphicx}
\usepackage{hhline}
\usepackage[multiple]{footmisc}
\usepackage{natbib}
\usepackage{url}
\usepackage{color}
\usepackage{multirow}
\usepackage{rotating} 
\usepackage{amsmath}
\usepackage{amssymb}
\usepackage{hyphenat}
\usepackage[onelanguage,ruled,vlined]{algorithm2e}
\usepackage{pgfplots}
\pgfplotsset{
tick label style = {font=\footnotesize},
every axis label/.append style={font=\footnotesize},
}
\pgfplotsset{compat=1.3}
\usetikzlibrary{patterns}
\usetikzlibrary{calc}
\usepackage{enumerate}

\journalname{Social Network Analysis and Mining}
\begin{document}

\title{Event detection, tracking and visualization in Twitter}
\subtitle{A mention-anomaly-based approach}


\author{Adrien Guille         \and
        C\'ecile Favre 
}


\institute{A. Guille \at
              ERIC Lab, University of Lyon 2\\ 
              5 avenue Pierre Mendès France \\
              69676 Bron, France\\
              \email{adrien.guille@univ-lyon2.fr}           
           \and
           C. Favre \at
              cecile.favre@univ-lyon2.fr
}

\date{Received: December 14, 2014 / Revised: April 3, 2015 / Accepted: May 13, 2015}

\maketitle

\begin{abstract}
The ever-growing number of people using Twitter makes it a valuable source of timely information. However, detecting events in Twitter is a difficult task, because tweets that report interesting events are overwhelmed by a large volume of tweets on unrelated topics. Existing methods focus on the textual content of tweets and ignore the social aspect of Twitter. In this paper we propose \textit{MABED} (mention\hyp{}anomaly\hyp{}based Event Detection), a novel statistical method that relies solely on tweets and leverages the creation frequency of dynamic links (\textit{i.e.} mentions) that users insert in tweets to detect significant events and estimate the magnitude of their impact over the crowd. \textit{MABED} also differs from the literature in that it dynamically estimates the period of time during which each event is discussed, rather than assuming a predefined fixed duration for all events. The experiments we conducted on both English and French Twitter data show that the mention\hyp{}anomaly\hyp{}based approach leads to more accurate event detection and improved robustness in presence of noisy Twitter content. Qualitatively speaking, we find that \textit{MABED} helps with the interpretation of detected events by providing clear textual descriptions and precise temporal descriptions. We also show how \textit{MABED} can help understanding users' interest. Furthermore, we describe three visualizations designed to favor an efficient exploration of the detected events.

\end{abstract}

\section{Introduction}

\noindent Twitter is a social networking and micro-blogging service that allows users to publish short messages limited to 140 characters, \textit{i.e.} tweets. Users share, discuss and forward various kinds of information -- ranging from personal daily events to important and global event related information -- in real-time. The ever-growing number of users around the world tweeting makes Twitter a valuable source of timely information. On the other hand, it gives rise to an information overload phenomenon and it becomes increasingly difficult to identify relevant information related to \textit{significant events}. An event is commonly defined as a thing that happens at one specific time \citep{becker_events,aggarwal_event-detection}, and it is significant if it may be discussed by traditional media \citep{mcminn_corpus}. These facts raise the following question: \textit{How can we use Twitter for automated significant event detection and tracking?} The answer to this question would help analyze which events, or types of events, most interest the crowd. This is critical to applications for journalistic analysis, playback of events, \textit{etc}. Yet the list of ``trends'' determined by Twitter isn't so helpful since it only lists isolated keywords and provides no information about the level of attention it receives from the crowd nor temporal indications.

Twitter delivers a continuous stream of tweets, thus allowing the study of how topics grow and fade over time \citep{yang_twitter7}. In particular, event detection methods focus on detecting ``bursty'' patterns -- which are intuitively assumed to signal events \citep{kleinberg_burst} -- using various approaches ranging from term-weighting-based approaches \citep{shamma_peaky-topics,benhardus_twitter-tfidf} to topic-modeling-based approaches \citep{lau_online-lda,hu_icwsm12}, including clustering-based approaches \citep{weng_edcow,chenliang_twevent,parikh_et}. Despite the wealth of research in the area, the vast majority of prior work focuses on the textual content of tweets and mostly neglects the social aspect of Twitter. However, users often insert extra-textual content in their tweets. Of particular interest is the ``mentioning practice'', which consists of citing other users' screennames in tweets (using the syntax ``@username''). Mentions are in fact dynamic links created either intentionally to engage the discussion with specific users or automatically when replying to someone or re-tweeting. This type of link is dynamic because it is related to a particular time period, \textit{i.e.} the tweet lifespan, and a particular topic, \textit{i.e.} the one being discussed. 

\noindent\textbf{Proposal}~~ We tackle the issue of event detection and tracking in Twitter by devising a new statistical method, named \textit{MABED} (mention\hyp{}anomaly\hyp{}based Event Detection). It relies solely on statistical measures computed from tweets and produces a list of events, each event being described by (i) a main word and a set of weighted related words, (ii) a period of time and (iii) the magnitude of its impact over the crowd. In contrast with existing methods, \textit{MABED} doesn't only focus on the textual content of tweets but also leverages the frequency at which users interact through mentions, with the aim to detect more accurately the most impactful events. It also differs from the literature in that it dynamically estimates the period of time during which each event is discussed, rather than assuming a predefined fixed duration for all events, in order to provide clearer event descriptions. What is more, we develop three interactive visualizations to ensure an efficient exploration of the detected events: (i) a timeline that allows exploring events through time, (ii) a chart that plots the magnitude of impact of events through time and (iii) a graph that allows identifying semantically related events. The implementation of \textit{MABED} is available for re-use and future research. It is also included in \textit{SONDY} \citep{guille_sondy}, an open-source social media data mining software that implements several state-of-the-art algorithms for event detection.

\noindent\textbf{Results}~~ We perform quantitative and qualitative studies of the proposed method on both English and French Twitter corpora containing respectively about 1.5 and 2 millions tweets. We show that \textit{MABED} is able to extract an accurate and meaningful retrospective view of the events discussed in each corpus, with short computation times. To study precision and recall, we ask human annotators to judge whether the detected events are meaningful and significant events. We demonstrate the relevance of the mention\hyp{}anomaly\hyp{}based approach, by showing that \textit{MABED} outperforms a variant that ignores the presence of mentions in tweets. We also show that \textit{MABED} advances the state-of-the-art by comparing its performance against those of two recent methods from the literature. The analysis of these results suggests that considering the frequency at which users interact through mentions leads to more accurate event detection and improved robustness in presence of noisy Twitter content. Lastly, we analyze the types of events detected by \textit{MABED} with regard to the communities detected in the network structure (\textit{i.e.} the following relationships) that interconnects the authors of the tweets. The results of this analysis shed light on the interplay between the social and topical structures in Twitter and show that \textit{MABED} can help understanding users' interests.

The rest of this paper is organized as follows. In the next section we discuss related work, before describing in detail the proposed method in Section \ref{sec:proposed-method}. Then an experimental study showing the method's effectiveness and efficiency is presented in Section \ref{sec:experiments}. Next, we present three visualizations for exploring the detected events. Finally, we conclude and discuss future work in Section~\ref{sec:conclusion}.

\section{Related Work}
\label{sec:related-work}

Methods for detecting events in Twitter rely on a rich body of work dealing with event, topic and burst detection from textual streams. In a seminal work, \cite{kleinberg_burst} studies time gaps between messages in order to detect bursts of email messages. Assuming that all messages are about the same topic, he proposes to model bursts with hidden Markov chains. \cite{alsumait_olda} propose \textit{OLDA} (On-line Latent Dirichlet Allocation), a dynamic topic model based on \textit{LDA} \citep{blei_lda}. It builds evolutionary matrices translating the evolution of topics detected in a textual stream through time, from which events can be identified. \cite{fung_vldb} propose to detect and then cluster bursty words by looking at where the frequency of each word in a given time window is positioned in the overall distribution of the number of documents containing that word.

Tweet streams differ from traditional textual document streams, in terms of publishing rate, content, \textit{etc}. Therefore, developing event detection methods adapted to Twitter has been studied in several papers in recent years. Next, we give a brief survey of the proposed approaches. 

\subsection{Event Detection and Tracking from Tweets}

\noindent\textbf{Term-weighting-based approaches}~~ The \textit{Peakiness Score} \citep{shamma_peaky-topics} is a normalized word frequency metric, similar to the $tf \cdot idf$ metric, for identifying words that are particular to a fixed length time window and not salient in others. However, individual words may not always be sufficient to describe complex events because of the possible ambiguity and the lack of context. To cope with this, \cite{benhardus_twitter-tfidf} propose a different normalized frequency metric, \textit{Trending Score}, for identifying event-related n-grams. For a given n-gram and time window, it consists in computing the normalized frequency, $tf_{\textit{norm}}$, of that n-gram with regard to the frequency of the other n-grams in this window. The \textit{Trending Score} of a n-gram in a particular time window is then obtained by normalizing the value of $tf_{\textit{norm}}$ in this time window with regard to the values calculated in the others.

\noindent\textbf{Topic-modeling-based approaches}~~ \cite{lau_online-lda} propose an online variation of \textit{LDA}. The idea is to incrementally update the topic model in each time window using the previously generated model to guide the learning of the new model. At every model update, the word distribution in topics evolves. Assuming that an event causes a sudden change in the word distribution of a topic, authors propose to detect events by monitoring the degree of evolution of topics using the Jensen-Shannon divergence measure. \cite{hu_icwsm12} note that topic modeling methods behave badly when applied to short documents such as tweets. To remedy this, they propose \textit{ET-LDA} (joint Event and Tweets \textit{LDA}). It expands tweets with the help of a search engine and then aligns them with re\hyp{}transcriptions of events provided by traditional media, which heavily influences the results. Globally, topic\hyp{}modeling\hyp{}based methods suffer from a lack of scalability, which renders their application to tweet streams difficult. However, works by \cite{aiello_sensing-trending-topics} reveal that dynamic topic models don't effectively handle social streams in which many events are reported in parallel.

\noindent\textbf{Clustering-based approaches}~~ \textit{EDCoW} \citep{weng_edcow} breaks down the frequency of single words into wavelets and leverages Fourier and Shannon theories to compute the change of wavelet entropy to identify bursts. Trivial words are filtered away based on their corresponding signal's auto correlation, and the similarity between each pair of non-trivial words is measured using cross correlation. Eventually, events are defined as bags of words with high cross correlation during a predefined fixed time window, detected with modularity\hyp{}based graph clustering. However, as pointed out by \cite{chenliang_twevent} and \cite{parikh_et}, measuring cross correlation is computationally expensive. Furthermore, measuring similarity utilizing only cross correlation can result in clustering together several unrelated events that happened in the same time span. \textit{TwEvent} \citep{chenliang_twevent} detects event from tweets by analyzing n-grams. It filters away trivial n-grams based on statistical information derived from Wikipedia and the Microsoft Web N-Gram service. The similarity between each pair of non-trivial n-grams is then measured based on frequency and content similarity, in order to avoid merging distinct events that happen concurrently. Then, similar n-grams in fixed\hyp{}length time windows are clustered together using a $k$-nearest neighbor strategy. Eventually, the detected events are filtered using, again, statistical information derived from Wikipedia. As a result, the events detected with \textit{TwEvent} are heavily influenced by Microsoft Web N-Gram and Wikipedia, which could potentially distort the perception of events by Twitter users and also give less importance to recent events that are not yet reported on Wikipedia. It is also worth mentioning \textit{ET} \citep{parikh_et}, a recent method similar to \textit{TwEvent}, except that it doesn't make use of external sources of information and focuses on bigrams. The similarity between pairs of bigrams is measured based on normalized frequency and content similarity, and the clustering is performed using a hierarchical agglomerative strategy.

\subsection{Event Visualization in Twitter}

\textit{Eddi} \citep{bernstein_eddi} is among the first tools developed for visualizing events from tweets. It displays a single word cloud that describes all the detected events, as well as a single stacked area chart that plots the evolution of the relative volume of tweets for each event. \cite{mathioudakis_twittermonitor} propose \textit{TwitterMonitor}, a system that allows for a finer understanding of the detected events in comparison with \textit{Eddi}. It displays a list of events, each event being described by a set of words and a chart that plots the evolution of the volume of related tweets. \textit{KeySEE} \citep{lee_keysee} is a tool that offers similar functionalities with more sophisticated visualizations, such as word clouds to describe events instead of sets of words. \cite{marcus_twitinfo} describe \textit{TwitInfo}, a tool whose interface revolves around a timeline of events. The user can click an event to see the tweets published during the related time interval, or to see the most cited URLs in these tweets. Let us also mention work by \cite{kraft_gtac}, in which a heatmap describes the distribution of events across time.

\section{Proposed Method}
\label{sec:proposed-method}

In this section, we first formulate the problem we intend to solve. Then we give an overview of the solution we propose, \textit{MABED}, before describing it formally.

\subsection{Problem Formulation}

\noindent\textbf{Input}~~ We are dealing with a tweet corpus $\mathcal{C}$. We discretize the time-axis by partitioning the tweets into $n$ time-slices of equal length. Let $V$ be the vocabulary of the words used in all the tweets and $V_{\emph{@}}$ be the vocabulary of the words used in the tweets that contain at least one mention. Table \ref{tab:notations} gives the definitions of the notations used in the rest of this paper.

\begin{table}[]
\centering
\caption{Table of notations.}
\begin{footnotesize}
\begin{tabular}{cl}
\hline 
Notation & Definition \\ 
\hline
$N$ & Total number of tweets in the corpus \\ \hline
$N^i$ & Number of tweets in the $i^{\text{th}}$ time-slice \\ \hline
\multirow{ 2}{*}{$N_t^i$} & Number of tweets in the $i^{\text{th}}$ time-slice \\ 
 & that contain the word $t$ \\\hline
\multirow{ 2}{*}{$N_{\emph{{@}}t}$} & Number of tweets in the corpus that \\
 & contain the word $t$ and at least one mention \\ \hline
\multirow{ 2}{*}{$N_{\emph{{@}}t}^i$} &  Number of tweets that contain the word $t$ \\
 & and at least one mention in the $i^{\text{th}}$ time-slice \\ 
\hline 
\end{tabular} 
\end{footnotesize}
\label{tab:notations}
\end{table}

\noindent\textbf{Output}~~ The objective is to produce a list $L$, such that $|L| = k$, containing the events with the $k$ highest magnitude of impact over the crowd's tweeting behavior. We define an event as a bursty topic, with the magnitude of its impact characterized by a score. Definitions \ref{def:bursty-topic} and \ref{def:event} below respectively define the concepts of bursty topic and event.

\begin{definition}[Bursty Topic]
Given a time interval $I$, a topic $T$ is considered bursty if it has attracted an uncommonly high level of attention (in terms of creation frequency of mentions) during this interval in comparison to the rest of the period of observation. The topic $T$ is defined by a main term $t$ and a set $S$ of weighted words describing it. Weights vary between 0 and 1. A weight close to 1 means that the word is central to the topic during the bursty interval whereas a weight closer to 0 means it is less specific.
\label{def:bursty-topic}
\end{definition}

\begin{definition}[Event]
An event $e$ is characterized by a bursty topic $\mathcal{BT} = [T,I]$ and a value $\textit{Mag} > 0$ indicating the magnitude of the impact of the event over the crowd.
\label{def:event}
\end{definition}

\subsection{Overview of the Proposed Method}

\begin{figure}[ht]
\begin{center}
\includegraphics[width=0.98\columnwidth]{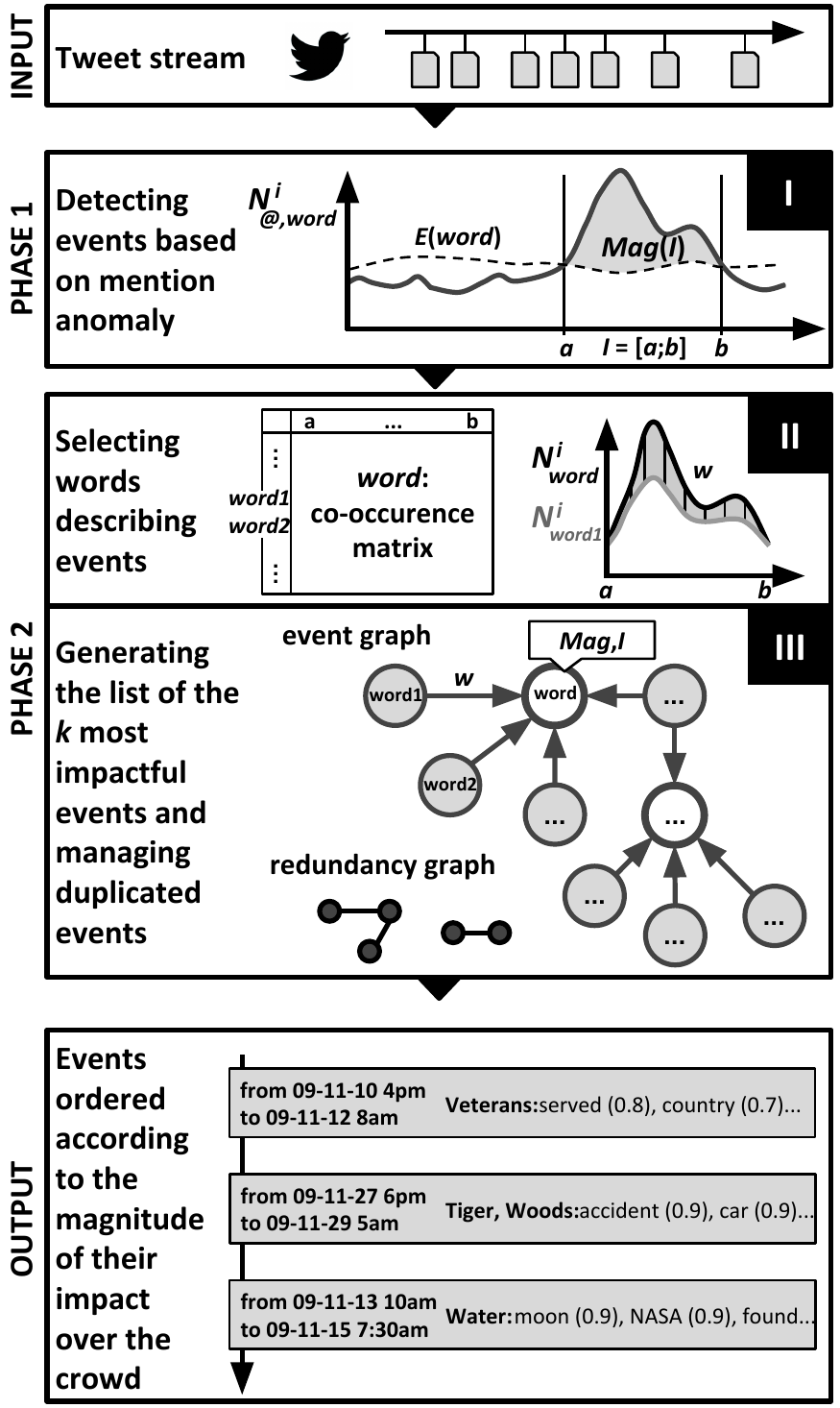}
\caption{Overall flow of the proposed method, \textit{MABED}.}
\label{fig:flow}
\end{center}
\end{figure}

The method has a two-phase flow. It relies on three components: (i) the detection of events based on mention\hyp{}anomaly, (ii) the selection of words that best describe each event and (iii) the generation of the list of the $k$ most impactful events. The overall flow, illustrated on Figure \ref{fig:flow}, is briefly described hereafter.

\begin{enumerate}
\item The mention creation frequency related to each word $t \in V_{\emph{@}}$ is analyzed with the first component. The result is a list of partially defined events, in that they are missing the set $S$ of related words. This list is ordered according to the impact of the events.
\item The list is iterated through starting from the most impactful event. For each event, the second component selects the set $S$ of words that best describe it. The selection relies on measures based on the co-occurrence and the temporal dynamics of words tweeted during $I$. Each event processed by this component is then passed to the third component, which is responsible for storing event descriptions and managing duplicated events. Eventually, when $k$ distinct events have been processed, the third component merges duplicated events and returns the list $L$ containing the top $k$ events.
\end{enumerate}


\subsection{Detection of Events Based on Mention Anomaly}

The objective of this component is to precisely identify when events happened and to estimate the magnitude of their impact over the crowd. It relies on the identification of bursts based on the computation of the anomaly in the frequency of mention creation for each individual word in $V_{\emph{{@}}}$. Existing methods usually assume a fixed duration for all events that corresponds to the length of a time-slice. It's not the case with \textit{MABED}. In the following, we describe how to compute the anomaly of a word for a given time-slice, then we describe how to measure the magnitude of impact of a word given a contiguous sequence of time-slices. Eventually, we show how to identify the intervals that maximize the magnitude of impact for each word in $V_{\emph{{@}}}$.

\noindent\textbf{Computation of the anomaly at a point}~~ Before formulating the anomaly measure, we define the expected number of mention creation associated to a word $t$ for each time-slice $i \in [1;n]$. We assume that the number of tweets that contain the word $t$ and at least one mention in the $i^{\text{th}}$ time-slice, $N_{\emph{{@}}t}^i$, follows a generative probabilistic model. Thus we can compute the probability $P(N_{\emph{{@}}t}^i)$ of observing $N_{\emph{{@}}t}^i$. For a large enough corpus, it seems reasonable to model this kind of probability with a binomial distribution \citep{fung_vldb}. Therefore we can write:
$$
P(N_{\emph{{@}}t}^i) = \binom{N^i}{N_{\emph{{@}}t}^i}p_{\emph{{@}}t}^{N_{\emph{{@}}t}^i}(1-p_{\emph{{@}}t})^{N^i-N_{\emph{@}t}^i}
$$
where $p_{\emph{{@}}t}$ is the expected probability of a tweet containing $t$ and at least one mention in any time-slice. Because $N^i$ is large we further assume that $P(N_{\emph{{@}}t}^i)$ can be approximated by a normal distribution \citep{chenliang_twevent}, that is to say:
$$
P(N_{\emph{{@}}t}^i) \sim \mathcal{N}(N^ip_{\emph{{@}}t},N^ip_{\emph{{@}}t}(1-p_{\emph{{@}}t}))
$$
It follows that the expected frequency of tweets containing the word $t$ and at least one mention in the $i^{\text{th}}$ time-slice is:
$$
E[t|i] = N^ip_{\emph{{@}}t} \text{ where } p_{\emph{{@}}t} = N_{\emph{{@}}t} / N
$$
Eventually, we define the anomaly of the mention creation frequency related to the word $t$ at the $i^{\text{th}}$ time-slice this way:
$$
\textit{anomaly}(t,i) = N_{\emph{{@}}t}^i - E[t|i]
$$
With this formulation, the anomaly is positive only if the observed mention creation frequency is strictly greater than the expectation. Event-related words that are specific to a given period of time are likely to have high anomaly values during this interval. In contrast, recurrent (\textit{i.e.} trivial) words that aren't event-specific are likely to show little discrepancy from expectation. What is more, as opposed to more sophisticated approaches like modeling frequencies with Gaussian mixture models, this formulation can easily scale to the number of distinct words used in tweets.

\noindent\textbf{Computation of the magnitude of impact}~~ The magnitude of impact, \textit{Mag}, of an event associated with the time interval $I=[a;b]$ and the main word $t$ is given by the formula below. It corresponds to the algebraic area of the anomaly function on $[a;b]$.
$$
\textit{Mag}(t,I) = \displaystyle \int_{a}^{b} \textit{anomaly}(t,i) \, \mathrm{d}i = \displaystyle \sum_{i=a}^{b} \textit{anomaly}(t,i)
$$
The algebraic area is obtained by integrating the discrete anomaly function, which in this case boils down to a sum. 

\begin{figure}[]
\begin{center}
\includegraphics[width=0.98\columnwidth]{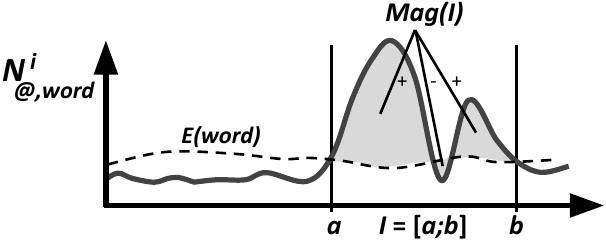}
\caption{Identification of the time interval that maximizes the magnitude of impact for a given word.}
\label{fig:phase1}
\end{center}
\end{figure}

\noindent\textbf{Identification of events}~~ For each word $t \in V_{\emph{@}}$, we identify the interval $I$ that maximizes the magnitude of impact, that is to say :
$$
I = \underset{I} {\mathrm{argmax}} ~\textit{Mag}(t,I)
$$
Because the magnitude of impact of an event described by the main word $t$ and the time interval $I$ is the sum of the anomaly measured for this word over $I$, this optimization problem is similar to a ``Maximum Contiguous Subsequence Sum'' (\textit{MCSS}) problem. The \textit{MCSS} problem is well known and finds application in many fields \citep{fan_max-sum-bioinformatics,lappas_discrepancy-model}. In other words, for a given word $t$ we want to identify the interval $I=[a;b]$, such that:
$$
\textit{Mag}(t,I) = \max \lbrace \sum_{i=a}^{b} \textit{anomaly}(t,i) | 1 \leqslant a \leqslant b \leqslant n\rbrace
$$
This formulation permits the anomaly to be negative at some points in the interval, as shown in Figure \ref{fig:phase1}, only if it permits extending the interval while increasing the total magnitude. This is a desirable property, as it avoids fragmenting events that last several days because of the lower activity on Twitter during the night for instance, which can lead to low or negative anomaly. Another desirable property of this formulation is that a given word can't be considered as the main word of more than one event. This increases the readability of events for the following reason. The bigger the number of events that can be described by a given word, the less specific to each event this word is. Therefore, this word should rather be considered as a related word than the main word. We solve this \textit{MCSS} type of problem using the linear-time algorithm described by \cite{bentley_kadane}. Eventually, each event detected following this process is described by: (i) a main word $t$ (ii) a period of time $I$ and (iii) the magnitude of its impact over the tweeting behavior of the users, $\textit{Mag}(t,I)$.

\subsection{Selection of Words Describing Events}

Observing that clustering\hyp{}based methods can in some cases lead to noisy event descriptions \citep{valkanas_twinsight}, we adopt a different approach which we describe hereafter, with the aim to provide more semantically meaningful descriptions.

In order to limit information overload, we choose to bound the number of words used to describe an event. This bound is a fixed parameter noted $p$. We justify this choice because of the shortness of tweets. Indeed, because tweets contain very few words, it doesn't seem reasonable for an event to be associated with too many words \citep{weng_edcow}.

\noindent\textbf{Identification of the candidate words}~~ The set of candidate words for describing an event is the set of the words with the $p$ highest co-occurrence counts with the main word $t$ during the period of time $I$. The most relevant words are selected amongst the candidates based on the similarity between their temporal dynamics and the dynamics of the main word during $I$. For that, we compute a weight $w_{q}$ for each candidate word $t'_q$. We propose to estimate this weight from the time\hyp{}series for $N_t^i$ and $N_{t'_q}^i$ with the correlation coefficient proposed by \cite{erdem_correlation}. This coefficient, primarily designed to analyze stock prices, has two desirable properties for our application: (i) it is parameter\hyp{}free and (ii) there is no stationarity assumption for the validity of this coefficient, contrary to common coefficients, \textit{e.g.} Pearson's coefficient. This coefficient takes into account the lag difference of data points in order to better capture the direction of the co-variation of the two time\hyp{}series over time. For the sake of conciseness, we directly give the formula for the approximation of the coefficient, given words $t$, $t'_q$ and the period of time $I = [a;b]$:
$$
\rho_{O t,t'_q}  = \frac{\displaystyle\sum_{i=a+1}^b A_{t,t'_q}}{(b-a-1)A_tA_{t'_q}}
$$
$$\text{where } A_{t,t'_q} = (N_t^i-N_t^{i-1}) (N_{t'_q}^i-N_{t'_q}^{i-1})\text{,}$$
$$A^2_t =  \frac{\sum_{i=a+1}^b(N_t^i-N_t^{i-1})^2}{b-a-1}\text{ and}$$
$$A^2_{t'_q} = \frac{\sum_{i=a+1}^b(N_{t'_q}^i-N_{t'_q}^{i-1})^2}{b-a-1}$$
This practically corresponds to the first order auto\hyp{}correlation of the time\hyp{}series for $N^i_t$ and $N^i_{t'_q}$. The proof that $\rho_O$ satisfies $\vert \rho_O \vert \leqslant 1$ using the Cauchy\hyp{}Schwartz inequality is given by \cite{erdem_correlation}. Eventually, we define the weight of the term $t'_q$ as an affine function of $\rho_O$ to conform with our definition of bursty topic, \textit{i.e.} $0 \leqslant w_q \leqslant 1$:
$$
w_q = \frac{\rho_{O t,t'_q}+1}{2}
$$
Because the temporal dynamics of very frequent words are less impacted by a particular event, this formulation -- much like $tf \cdot idf$ -- diminishes the weight of words that occur very frequently in the stream and increases the weight of words that occur less frequently, \textit{i.e.} more specific words.

\noindent\textbf{Selection of the most relevant words}~~ The final set of words retained to describe an event is the set $S$, such that $\forall t'_q \in S$, $w_q \geqslant \theta$. The parameters $p$ and $\theta$ allow the users of \textit{MABED} to adjust the level of information and detail they require.

\subsection{Generating the List of the Top k Events}

Each time an event has been processed by the second component, it is passed to the third component. It is responsible for storing the description of the events while managing duplicated events. For that, it uses two graph structures: the event graph and the redundancy graph. The first is a directed, weighted, labeled graph that stores the descriptions of the detected events. The representation of an event $e$ in this graph is as follows. One node represents the main word $t$ and is labeled with the interval $I$ and the score $\textit{Mag}$. Each related word $t'_q$ is represented by a node and has an arc toward the main word, which weight is $w_q$. The second structure is a simple undirected graph that is used to represent the relations between the eventual duplicated events, represented by their main words. 

\begin{figure}[]
\centering
\includegraphics[width=0.66\columnwidth]{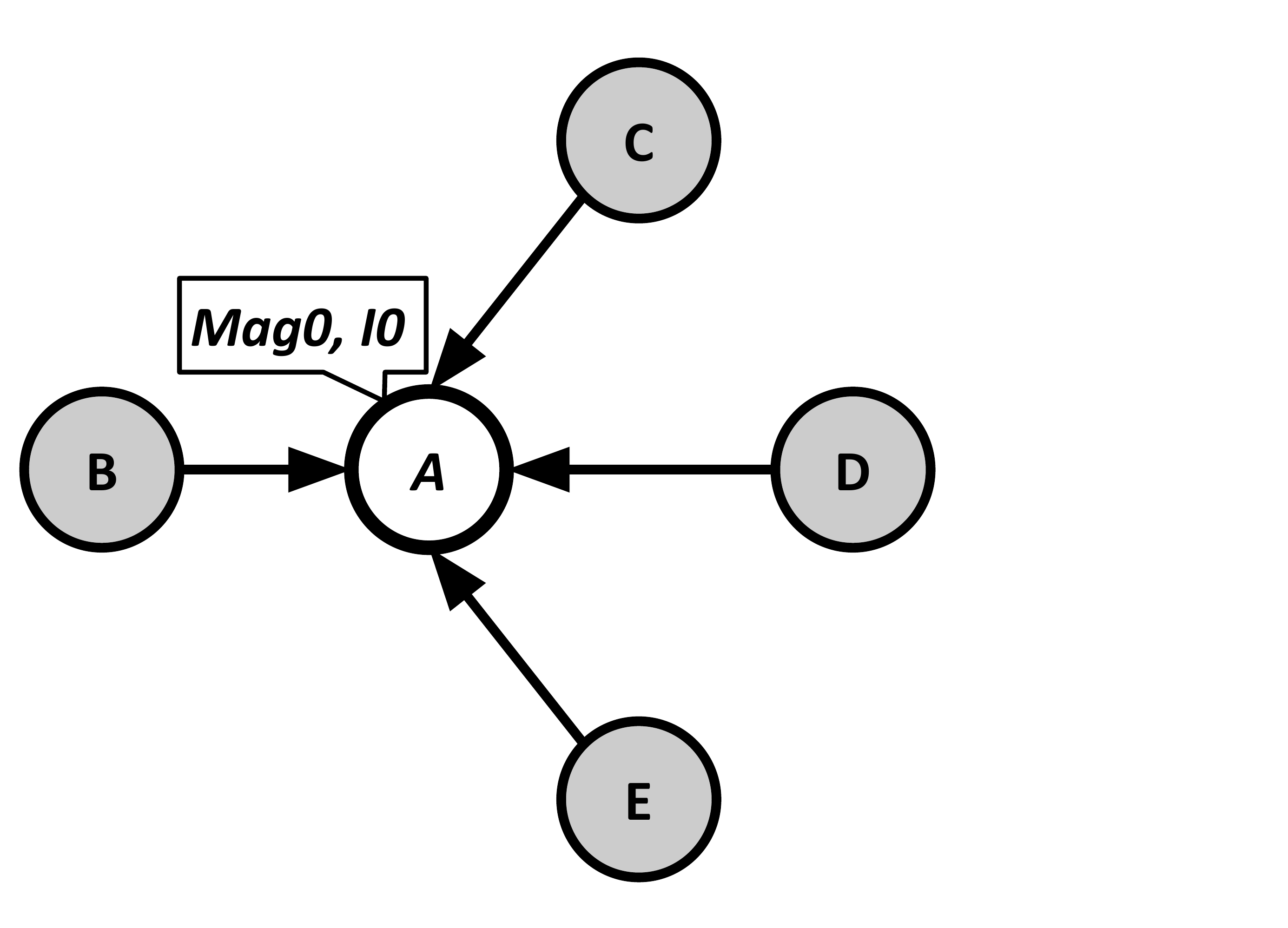}

Storing the description of $e_0$ in the event graph.

\includegraphics[width=0.66\columnwidth]{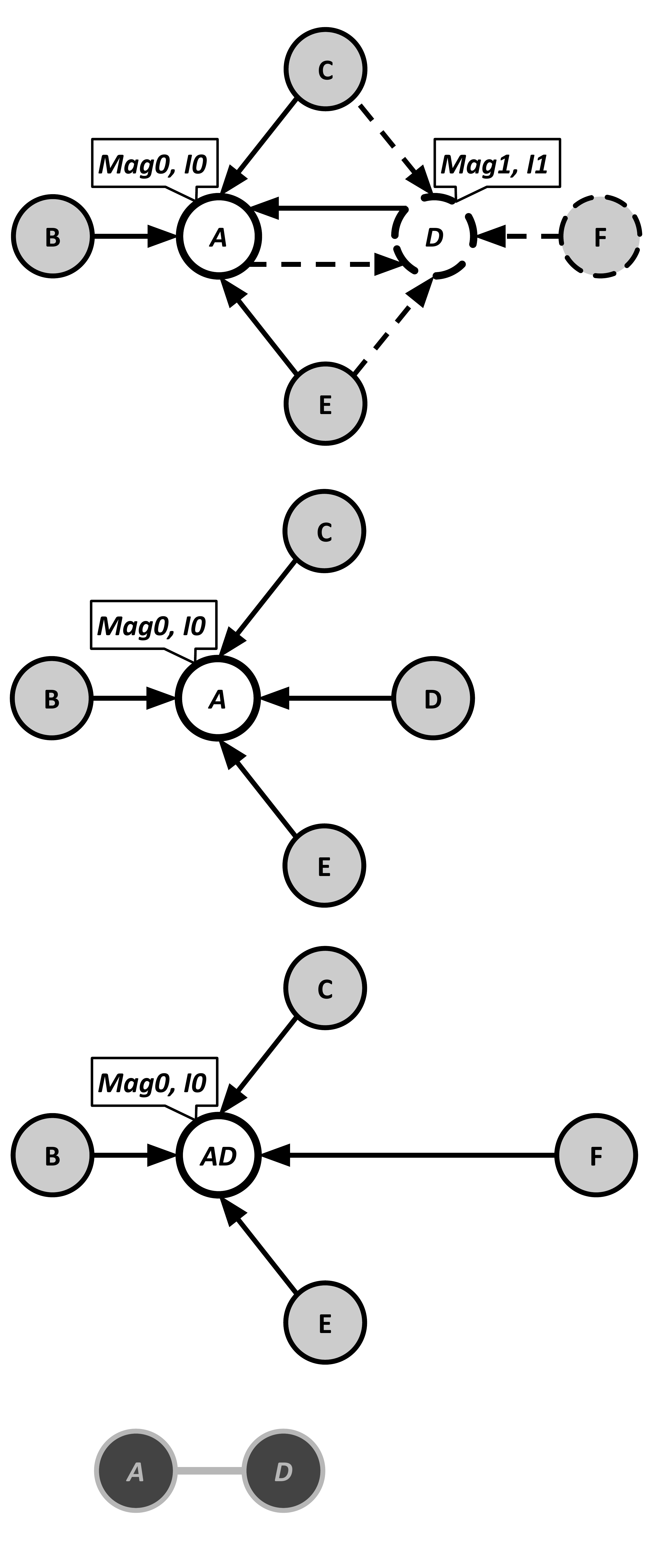}

Checking connectivity between events $e_0$ and $e_1$.

\includegraphics[width=0.66\columnwidth]{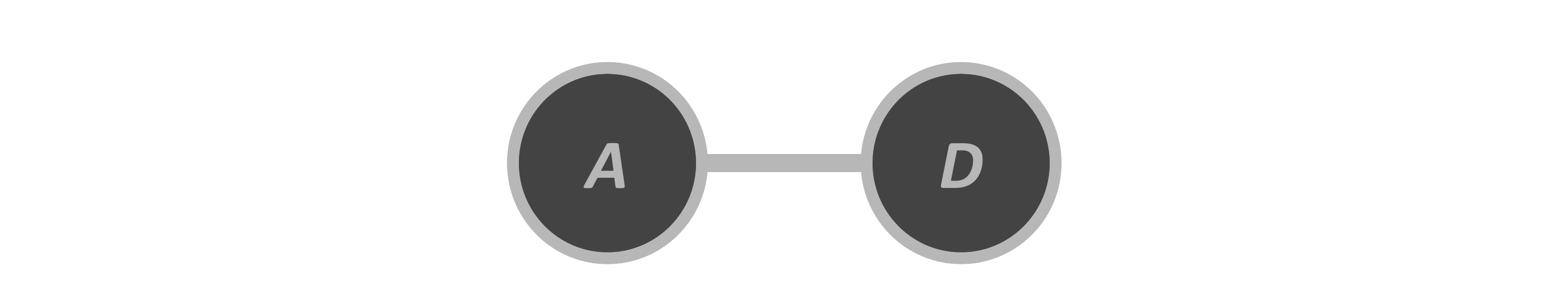}

Adding a link between \textit{A} and \textit{D} in the redundancy graph.

\caption{Detecting duplication between events $e_0$ and $e_1$.}
\label{fig:duplication}
\end{figure}

Let $e_1$ be the event that the component is processing. First, it checks whether it is a duplicate of an event that is already stored in the event graph or not. If it isn't the case, the event is added to the graph and the count of distinct events is incremented by one. Otherwise, assuming $e_1$ is a duplicate of the event $e_0$, a relation is added between $t_0$ and $t_1$ in the redundancy graph. When the count of distinct events reaches $k$, the duplicated events are merged and the list of the top $k$ most impactful events is returned. We describe how duplicated events are identified and how they are merged hereafter.

\begin{figure}[]
\centering
\includegraphics[width=0.66\columnwidth]{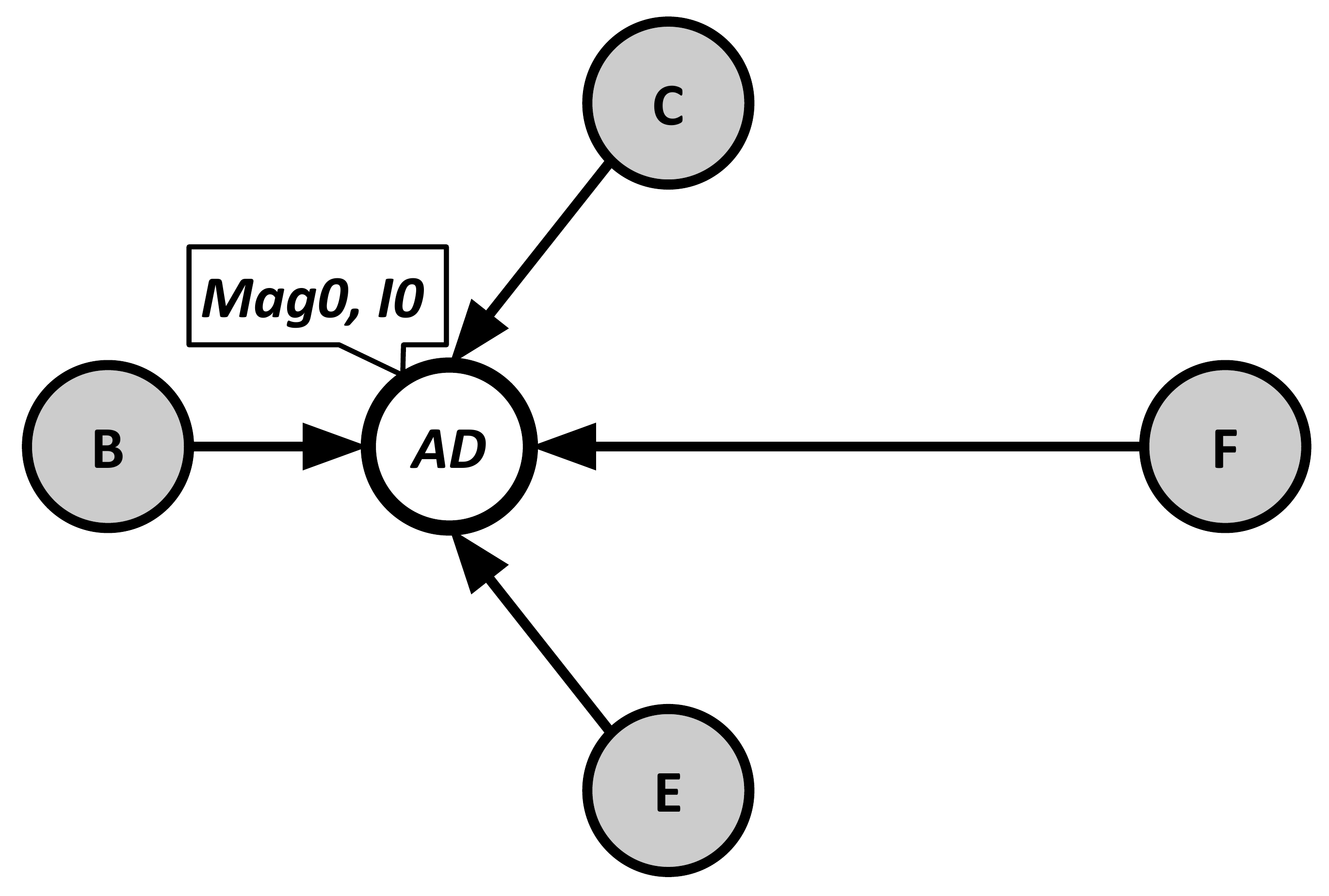}
\caption{Merging event $e_0$ with duplicated event $e_1$.}
\label{fig:merging}
\end{figure}

\noindent\textbf{Detecting duplicated events}~~ The event $e_1$ is considered to be a duplicate of the event $e_0$ already stored in the event graph if (i) the main words $t_1$ and $t_0$ would be mutually connected and (ii) if the overlap coefficient between the periods of time $I_1$ and $I_0$ exceeds a fixed threshold. The overlap coefficient is defined as $\frac{|I_1 \cap I_0|}{\min(I_1,I_0)}$ and the threshold is noted $\sigma$, $\sigma \in ]0;1]$. In this case, the description of $e_1$ is stored aside and a relation is added between $t_1$ and $t_0$ in the redundancy graph. An example is shown in Figure \ref{fig:duplication}, where $e_0 =$ $\{A$, $\{B,C,D,E\}$, $I_0$, $\textit{Mag}_{0}\}$, and $e_1 =$ $\{D$, $\{C,E,F\}$, $I_1$, $\textit{Mag}_{1}\}$.

\noindent\textbf{Merging duplicated events}~~ Identifying which duplicated events should be merged together is equivalent to identifying the connected components in the redundancy graph. This is done in linear time, w.r.t to the numbers vertices and edges of the graph, using the algorithm described by \cite{hopcroft_connected-components}. In each connected component, there is exactly one node that corresponds to an event stored in the event graph. Its magnitude of impact and the related time interval remain the same, but its textual description is updated according to the following principle. The main word becomes the aggregation of the main words of all duplicated events. The words describing the updated event are the $p$ words among all the words describing the duplicated events with the $p$ highest weights. Figure \ref{fig:merging} shows the description of the event resulting from the merging of the events $e_0$ and $e_1$, based on the event and redundancy graphs shown in Figure \ref{fig:duplication}.

\subsection{Overall algorithm}

To conclude this section, Algorithm 1 sums-up the overall flow of \textit{MABED}.

\begin{algorithm}[]
 \SetAlgoLined
 \KwData{A corpus $\mathcal{C}$ of $N$ tweets partitioned into $n$ time-slices of equal length and the corresponding vocabularies, $V$ and $V_{\emph{@}}$}
 \textbf{Parameters}: $k > 0$, $p > 0$, $\theta \in [0;1]$ and $\sigma \in ]0;1]$\\
 \KwResult{The ordered list $L$ of the $k$ most impactful detected events}
 \tcc{First phase}
 Initialize the stack $P$ which stores the detected events during the first phase\;
 \For{each word $t \in V_{\emph{@}}$}{
     Identify the interval $I = [a;b]$ such that : 
     $\textit{Mag}(t,I) = \max \lbrace \sum_{i=a}^{b} \textit{anomalie}(t,i)$\;
     Add the event $e = [t,\emptyset,I,\textit{Mag}(t,I)]$ in $P$\; 
 }
 Sort the event stack $P$ by descending magnitude of impact\;
 \tcc{Second phase}
 Initialize the event graph $G_E$ and the redundancy graph $G_R$\;
 Set \textit{count} to 0\;
 \While{$\textit{count} < k$ and $|P| > 0$}{
 Pop the event $e$ on top of the stack $P$\;
 Select the words that describe $e$, with parameters $p$ and $\theta$\;
     \eIf{$e$ is redundant with the event $e'$ which is already in $G_E$, for a given $\sigma$}{
      Add a link between the mains word of events $e$ and $e'$ in $G_R$\;
      Store the description of $e$ aside\;
     }{
      Insert the description $e$ in graph $G_E$\;
      Increment \textit{count}\;
     }
 }
 Identify which events should be merged based on $G_R$ then update $G_E$\;
 Transform graph $G_E$ into list $L$\;
 Sort $L$ \;
 \textbf{return} $L$\;
 
\caption{Overall algorithm for \textit{MABED}.}
\label{algo:mabed}
\end{algorithm}

\section{Experiments}
\label{sec:experiments}

In this section we present the main results of the extensive experimental study we conducted on both English and French Twitter data to evaluate \textit{MABED}. In the quantitative evaluation, we demonstrate the relevance of the mention\hyp{}anomaly\hyp{}based approach and we quantify the performance of \textit{MABED} by comparing it to state\hyp{}of\hyp{}the\hyp{}art methods. To evaluate precision and recall, we ask human annotators to judge whether the detected events are meaningful and significant. In the qualitative evaluation, we show that the descriptions of the events detected by \textit{MABED} are semantically and temporally more meaningful than the descriptions provided by existing methods, which favors an easy understanding of the results. Lastly, we analyse the detected event with regard to user communities and find that \textit{MABED} can help understanding their interests.

\subsection{Experimental Setup}

\noindent\textbf{Corpora}~~ Since the Twitter corpora used in prior work aren't available we base our experiments on two different corpora. The first corpus -- noted $\mathcal{C}_{en}$ -- contains 1,437,126 tweets written in English, collected with a user-centric strategy. They correspond to all the tweets published in November 2009 by 52,494 U.S.-based users \citep{yang_twitter7}. This corpus contains a lot of noise and chatter. According to the study conducted by \cite{pearanalytics_twitter}, the proportion of non\hyp{}event\hyp{}related tweets could be as high as 50\%. The second corpus -- noted $\mathcal{C}_{fr}$ -- contains 2,086,136 tweets written in French, collected with a keyword-based strategy. We have collected these tweets in March 2012, during the campaign for the 2012 French presidential elections, using the Twitter streaming API with a query consisting of the names of the main candidates running for president. This corpus is focused on French politics. Trivial words are removed from both corpora based on English and French standard stop-word lists. All timestamps are in UTC. Table \ref{tab:datasets} gives further details about each corpus.

\begin{table}[]
\centering
\caption{Corpus Statistics. \emph{@}: proportion of tweets that contain mentions, \textit{RT}: proportion of retweets.}
\begin{footnotesize}
\begin{tabular}{cccccccc}
 \hline 
 Corpus & Tweets & Authors & \emph{@} & \textit{RT}\\ 
 \hline 
 $\mathcal{C}_{en}$ & 1,437,126 & 52,494 & 0.54 & 0.17 \\ 
 $\mathcal{C}_{fr}$ & 2,086,136 & 150,209 & 0.68 & 0.43 \\ 
 \hline 
 \end{tabular}
\end{footnotesize}
 \label{tab:datasets}  
\end{table}

\noindent\textbf{Baselines for comparison}~~ We consider two recent methods from the literature: \textit{ET} (clustering\hyp{}based) and \textit{TS} (term\hyp{}weighting\hyp{}based). \textit{ET} is based on the hierarchical clustering of bigrams using content and appearance patterns similarity \citep{parikh_et}. \textit{TS} is a normalized frequency metric for identifying n-grams that are related to events \citep{benhardus_twitter-tfidf}. We apply it to both bigrams (\textit{TS2}) and trigrams (\textit{TS3}). We also consider a variant of \textit{MABED}, noted $\alpha$-\textit{MABED}, that ignores the presence of mentions in tweets. This means that the first component detects events and estimates their magnitude of impact based on the values of $N_t^i$ instead of $N_{\emph{@}t}^i$. The reasoning for excluding a comparison against topic-modeling-based methods is that in preliminary experiments we found that they performed poorly and their computation times were prohibitive.

\noindent\textbf{Parameter setting}~~  For \textit{MABED} and $\alpha$-\textit{MABED}, we partition both corpora using 30 minute time-slices, which allows for a good temporal precision while keeping the number of tweets in each time-slice large enough. The maximum number of words describing each event, $p$, and the weight threshold for selecting relevant words, $\theta$, are parameters that allow the user to define the required level of detail. Given that the average number of words per sentence on Twitter is 10.7 according to the study conducted by \cite{OUP_tweets}, we fix $p$ to $10$. For the purpose of the evaluation, we set $\theta=0.7$ so judges are only presented with words that are closely related to each event. There is a parameter that can affect the performance of \textit{MABED}: $\sigma$. In the following, we report results for $\sigma = 0.5$ (we discuss the impact of $\sigma$ in Section \ref{sec:impact_sigma}).

For \textit{ET} and \textit{TS}, because they assume a fixed duration for all events -- which corresponds to the length of one time-slice -- we partition both corpora using 1-day time-slices like in prior work. \textit{ET} has two parameters, for which we use optimal values provided by the authors.

\noindent\textbf{Evaluation metrics}~~ The corpora don't come with ground truth, therefore we have asked two human annotators to judge whether the detected events are meaningful and significant. The annotators are French graduate students who aren't involved in this project. Their task consisted in reading the descriptions of the events detected by each method, and independently assign a rate to each description. This rate can be either 1, if the annotator decides that the description is meaningful and related to a significant event (\textit{i.e.} an event that may be covered in traditional media), or 0 in any other cases. They also had to identify descriptions similar to ones which had previously been rated, in order to keep track of duplicates. Considering that annotating events is a time consuming task for the annotators, we limit the evaluation to the 40 most impactful events detected by each method (\textit{i.e.} $k = 40$) in each corpus. We measure precision as the ratio of the number of detected events that both annotators have rated 1, which we refer to as $k'$, to the total number of detected events, $k$:
%
$$
P = \frac{k'}{k}
$$
%
Based on the number of duplicated events, $k''$, we define recall as the fraction of distinct significant events among all the detected events \citep{chenliang_twevent}: 
$$
R = \frac{k' - k''}{k}
$$
%
We also measure the DERate \citep{chenliang_twevent}, which denotes the percentage of events that are duplicates among all the significant events detected, that is to say $\text{DERate} = k''/k'$.

\begingroup\makeatletter\def\f@size{8.3}\check@mathfonts

\begin{table*}[]
\centering
\caption{Performance of the five methods on the two corpora.}
\begin{footnotesize}
\begin{tabular}{l|cccc|cccc}
\hline
 & \multicolumn{4}{c|}{Corpus: $\mathcal{C}_{en}$} & \multicolumn{4}{c}{Corpus: $\mathcal{C}_{fr}$}\\ Method & Precision & F-measure & DERate & Running-time & Precision & F-measure & DERate & Running-time \\ \hline
\textit{MABED} & 0.775 & 0.682 & 0.167 & 96s & 0.825 & 0.825 & 0 & 88s \\
$\alpha$-\textit{MABED} & 0.625 & 0.571 & 0.160 & 126s & 0.725 & 0.712 & 0.025 & 113s \\
\textit{ET} & 0.575 & 0.575 & 0 & 3480s & 0.700 & 0.674 & 0.071 & 4620s \\
\textit{TS2} & 0.600 & 0.514 & 0.250 & 80s & 0.725 & 0.671 & 0.138 & 69s \\\textit{TS3} & 0.375 & 0.281 & 0.4 & 82s & 0.700 & 0.616 & 0.214 & 74s \\
\hline
\end{tabular}
\end{footnotesize}
\label{tab:performance}  
\end{table*}

\subsection{Quantitative Evaluation}

Hereafter, we discuss the performance of the five considered methods, based on the rates assigned by the annotators. The inter-annotator agreement, measured with Cohen's Kappa \citep{landis_kappa}, is $\kappa \simeq 0.76$, showing a strong agreement.
Table \ref{tab:performance} (page \pageref{tab:performance}) reports the precision, the F-measure defined as the harmonic mean of precision and recall (\textit{i.e.} $2 \cdot \frac{P \cdot R}{P + R} $), the DERate and the running-time of each method for both corpora.

\noindent\textbf{Comparison against baselines}~~ We notice that \textit{MABED} achieves the best performance on the two corpora, with a precision of 0.775 and F-measure of 0.682 on $\mathcal{C}_{en}$, and a precision and a F-measure of 0.825 on $\mathcal{C}_{fr}$. Although \textit{ET} yields a better \textit{DERate} on $\mathcal{C}_{en}$, it still achieves lower precision and recall than \textit{MABED} on both corpora. Furthermore, we measure an average relative gain of 17.2\% over $\alpha$\hyp{}\textit{MABED} in the F-measure, which suggests that considering the mentioning behavior of users leads to more accurate detection of significant events in Twitter. Interestingly, we notice that \textit{MABED} outperforms all baselines in the F-measure with a bigger margin on $\mathcal{C}_{en}$, which contains a lot more noise than $\mathcal{C}_{fr}$ -- with up to 50\% non-event-related tweets according to the study conducted by \cite{pearanalytics_twitter}. This suggests that considering the mentioning behavior of users also leads to more robust detection of events from noisy Twitter content. The DERate reveals that none of the significant events detected in $\mathcal{C}_{fr}$ by \textit{MABED} were duplicated, whereas 6 of the significant events detected in $\mathcal{C}_{en}$ are duplicates. Furthermore, we find that the set of events detected by the four baseline methods is a sub-set of the events detected by \textit{MABED}. Further analysis of the results produced by $\alpha$\hyp{}\textit{MABED}, \textit{TS2} and \textit{TS3} reveals that most of non-significant events they detected are related to spam. The fact that most of these irrelevant events aren't detected by \textit{MABED} suggest that considering the presence of mentions in tweets helps filtering away spam. Concerning \textit{ET}, the average event description is 17.25 bigrams long (\textit{i.e.} more than 30 words). As a consequence, the descriptions contain some unrelated words. Specifically, irrelevant events are mostly sets of unrelated words that don't make any sense. This is due in part to the fact that clustering-based approaches are prone to aggressively grouping terms together, as \cite{valkanas_twinsight} stated in a previous study.

\endgroup

\pgfplotsset{height=0.53\columnwidth, width=1.00\columnwidth}
\begin{figure}[]
\centering
	\begin{tikzpicture}[]
		\begin{axis}[grid=major, legend style={anchor=north west, legend cell align=left, at={(0,1)}}, xtick \empty, xmin=0.4,xmax=1,xlabel={Subsample size},ylabel={Normalized runtime},extra x ticks={0.4,0.55,0.7,0.85,1}]
			\addplot+[mark=o, color=orange, line width=1pt] coordinates {
			(1,1)
            (0.85,0.81)
            (0.7,0.63)
            (0.55,0.43)
            (0.4,0.32)
			};
			\addlegendentry{{\footnotesize \textit{MABED}}}
			\addplot+[mark=square, color= darkgray, line width=1pt] coordinates {
			(1,0.34)
            (0.85,0.27)
            (0.7,0.22)
            (0.55,0.15)
            (0.4,0.10)
            };
			\addlegendentry{{\footnotesize \textit{MABED} (8 threads)}}
	\end{axis}
	\end{tikzpicture}
	\caption{Runtime comparison versus subsample size.}
	\label{fig:runtime}
\end{figure}
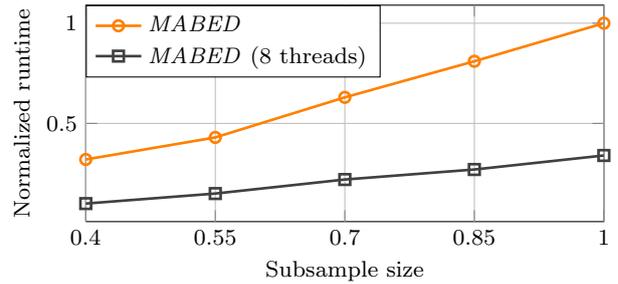

\pgfplotsset{height=0.66\columnwidth, width=1.00\columnwidth}
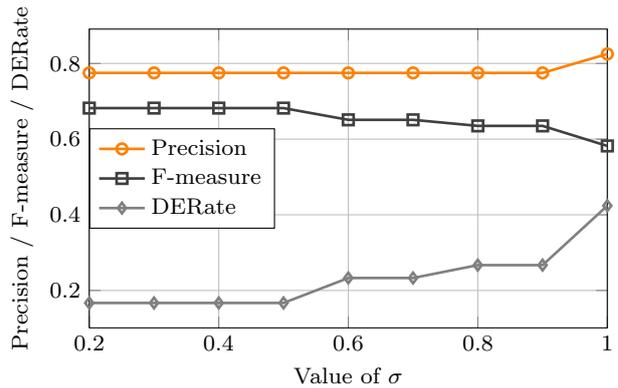
\begin{figure}[]
\centering
	\begin{tikzpicture}[]
		\begin{axis}[grid=major, legend style={anchor=west,legend cell align=left, at={(0,0.5)}},
    xlabel={Value of $\sigma$},
    ylabel={Precision / F-measure / DERate},
    xmin = 0.2,
    xmax = 1]
			\addplot+[mark=o, color=orange, line width=1pt] coordinates { 
			(0.2,0.775)
			(0.3,0.775)
			(0.4,0.775)
			(0.5,0.775)
			(0.6,0.775)
			(0.7,0.775)
			(0.8,0.775)
			(0.9,0.775)
			(1.0,0.825)
            };
			\addlegendentry{{\footnotesize Precision}}
			\addplot+[mark=square, color= darkgray, line width=1pt] coordinates {
			(0.2,0.682)
			(0.3,0.682)
			(0.4,0.682)
			(0.5,0.682)
			(0.6,0.651)
			(0.7,0.651)
			(0.8,0.635)
			(0.9,0.635)
			(1.0,0.582)
			};
			\addlegendentry{{\footnotesize F-measure}}
			\addplot+[mark=diamond, color= gray, line width=1pt] coordinates {
			(0.2,0.167)
			(0.3,0.167)
			(0.4,0.167)
			(0.5,0.167)
			(0.6,0.233)
			(0.7,0.233)
			(0.8,0.267)
			(0.9,0.267)
			(1.0,0.424)
			};
			\addlegendentry{{\footnotesize DERate}}
	\end{axis}
	\end{tikzpicture}
	\caption{Precision, F-measure and DERate of \textit{MABED} on $\mathcal{C}_{en}$ for different values of $\sigma$.}
	\label{fig:impact_sigma}
\end{figure}

\noindent\textbf{Efficiency}~~ It appears that \textit{MABED} and \textit{TS} have running\hyp{}times of the same order, whereas \textit{ET} is orders of magnitude slower, which is due to the clustering step that requires computing temporal and semantical similarity between all bigrams. We also observe that \textit{MABED} runs faster than $\alpha$-\textit{MABED}. The main reason for this is that $|V_\emph{{@}}| \leqslant |V|$, which speeds up the first phase. It should be noted that the running-times given in Table \ref{tab:performance} don't include the time required for preparing vocabularies and pre-computing term frequencies, which is more important for methods that rely on bigrams or trigrams. We evaluate the scalability of (i) \textit{MABED} and (ii) a parallelized version of \textit{MABED} (8 threads), by measuring their running-times on random subsamples of both corpora, for subsample sizes varying from 40\% to 100\%. Figure \ref{fig:runtime} shows the average normalized running-\hyp{}time versus subsample size. This means that the runtimes measured on a corpus are normalized by the longest runtime on this corpus and are then averaged for \textit{MABED} and \textit{MABED} (8 threads). We notice that runtimes grow linearly in size of the subsample. Furthermore, we note that \textit{MABED} (8 threads) is on average 67\% faster than \textit{MABED}.

\noindent\textbf{Impact of $\boldsymbol\sigma$ on \textit{MABED}}~~ 
\label{sec:impact_sigma}
While the list of events is constructed by \textit{MABED}, the overlap threshold $\sigma$ controls the sensitivity to duplicated events. Figure \ref{fig:impact_sigma} plots the precision, F-measure and DERate of \textit{MABED} on $\mathcal{C}_{en}$ for values of $\sigma$ ranging from 0.2 to 1. We observe that the value of $\sigma$ mainly impacts the DERate. More specifically, the DERate increases along the increase of $\sigma$ as fewer duplicated events are merged. For $\sigma = 1$, the precision increases to 0.825 because of the high percentage of duplicated significant events. Globally, it appears that the highest F-measure is attained for values of $\sigma$ ranging from 0.2 to 0.5. However, even using $\sigma = 1$, \textit{MABED} achieves a F-measure of 0.582, which is higher than all baselines on $\mathcal{C}_{en}$.

\subsection{Qualitative Evaluation}

\begin{table*}[ht]
\centering
\caption{Top 25 events with highest magnitude of impact over the crowd, detected by \textit{MABED} in $\mathcal{C}_{en}$. Main words are in bold and time intervals are given in UTC (Coordinated Universal Time).}
\begin{footnotesize}
\begin{tabular}{ccl}
\hline
$e_{\#}$ & Time interval & Topic\\ \hline 
\multirow{ 2}{*}{1} & from 25 09:30  & \textbf{thanksgiving, turkey}:  hope (0.72), happy (0.71)\\ 
 & to 28 06:30 & \textit{Twitter users celebrated Thanksgiving} \\
\hline
\multirow{ 2}{*}{2} & from 25 09:30   & \textbf{thankful}:  happy (0.77), thanksgiving (0.71)\\ 
 & to 27 09:00 & \textit{Related to event \#1}\\
\hline
\multirow{ 2}{*}{3} & from 10 16:00   & \textbf{veterans}:  served (0.80), country (0.78), military (0.73), happy (0.72)\\
 & to 12 08:00 & \textit{Twitter users celebrated the Veterans Day that honors people who have served in the U.S. army}\\
 \hline
\multirow{ 2}{*}{4} & from 26 13:00   & \textbf{black}:  friday (0.95), amazon (0.75)\\
 & to 28 10:30 & \textit{Twitter users were talking about the deals offered by Amazon the day before the ``Black Friday''} \\
 \hline
\multirow{ 2}{*}{5} & from 07 13:30  & \textbf{hcr, bill, health, house, vote}:  reform (0.92), passed (0.91), passes (0.88)\\
 & to 09 04:30 & \textit{The House of Representatives passed the health care reform bill on November 7} \\ \hline
\multirow{ 2}{*}{6} & from 05 19:30   & \textbf{hood, fort}:  ft (0.92), shooting (0.83), news (0.78), army (0.75), forthood (0.73)\\ 
 & to 08 09:00 & \textit{The Fort Hood shooting was a mass murder that took place in a U.S. military post on November 5}\\
\hline
\multirow{ 2}{*}{7} & from 19 04:30  & \textbf{chrome}:  os (0.95), google (0.87), desktop (0.71)\\ 
 & to 21 02:30 & \textit{On November 19, Google released Chrome OS's source code for desktop PC}\\
\hline
\multirow{ 2}{*}{8} & from 27 18:00 & \textbf{tiger, woods}:  accident (0.91), car (0.88), crash (0.88), injured (0.80), seriously (0.80)\\ 
 & to 29 05:00 & \textit{Tiger Woods was injured in a car accident on November 27, 2009}\\
\hline
\multirow{ 2}{*}{9} & from 28 22:30 & \textbf{tweetie, 2.1, app}:  retweets (0.93), store (0.90), native (0.89), geotagging (0.88)\\ 
 & to 30 23:30 & \textit{The iPhone app named Tweetie (v2.1), hit the app store with additions like retweets and geotagging} \\
\hline
\multirow{ 2}{*}{10} & from 29 17:00 & \textbf{monday, cyber}:  deals (0.84), pro (0.75)\\
 & to 30 23:30 & \textit{Twitter users were talking about the deals offered by online shops for the ``Cyber Monday''}\\
 \hline
\multirow{ 2}{*}{11} & from 10 01:00  & \textbf{linkedin}:  synced (0.86), updates (0.84), status (0.83), twitter (0.71)\\ 
& to 12 03:00 & \textit{Starting from November 10, LinkedIn status updates can be synced with Twitter}\\
\hline
\multirow{ 2}{*}{12} & from 04 17:00   & \textbf{yankees, series}:  win (0.84), won (0.84), fans (0.78), phillies (0.73), york (0.72)\\ 
 & to 06 05:30 & \textit{The Yankees baseball team defeated the Phillies to win their 27th World Series on November 4}\\
\hline
\multirow{ 2}{*}{13} & from 15 09:00  & \textbf{obama}:  chinese (0.75), barack (0.72), twitter (0.72), china (0.70)\\ 
 &  to 17 23:30 & \textit{Barack Obama admitted that he'd never used Twitter but Chinese should be able to}\\
\hline
\multirow{ 2}{*}{14} & from 25 10:00   & \textbf{holiday}:  shopping (0.72)\\ 
 & to 26 10:00 & \textit{Twitter users started talking about the ``Black Friday'', a shopping day and holiday in some states}\\
\hline
\multirow{ 2}{*}{15} & from 19 21:30   & \textbf{oprah, end}:  talk (0.81), show (0.79), 2011 (0.73), winfrey (0.71)\\ 
 & to 21 16:00 & \textit{On November 19, Oprah Winfrey announced her talk show will end in September 2011}\\
\hline
\multirow{ 2}{*}{16} & from 07 11:30   & \textbf{healthcare, reform}:  house (0.91), bill (0.88), passes (0.83), vote (0.83), passed (0.82)\\ 
 & to 09 05:00 & \textit{Related to event \#5}\\
\hline
\multirow{ 2}{*}{17} & from 11 03:30   & \textbf{facebook}:  app (0.74), twitter (0.73)\\ 
 & to 13 08:30 & \textit{No clear corresponding event}\\
\hline
\multirow{ 2}{*}{18} & from 18 14:00   & \textbf{whats}:  happening (0.76), twitter (0.73)\\
 & to 21 03:00 & \textit{Twitter started asking "What's happening?" instead of "What are you doing?" from November 18}\\
 \hline
\multirow{ 2}{*}{19} & from 20 10:00   & \textbf{cern}:  lhc (0.86), beam (0.79)\\ 
 & to 22 00:00 & \textit{On November 20, proton beams were successfully circulated in the ring of the LHC (CERN)}\\
\hline
\multirow{ 2}{*}{20} & from 26 08:00  & \textbf{icom}:  lisbon (0.99), roundtable (0.98), national (0.88)\\ 
 &  to 26 15:30 & \textit{The I-COM roundtable about market issues in Portugal took place on November 26}\\
 
\hline
\multirow{ 2}{*}{21} & from 03 23:00 & \textbf{maine}: voters (0.76), marriage (0.71)  \\ 
 &  to 05 10:00 & \textit{On November 4, Maine voters repealed a state law granting same-sex couples the right to marry}\\ 
 
\hline
\multirow{ 2}{*}{22} & from 07 13:00 & \textbf{droid}: verizon (0.75), iphone (0.72), video (0.70) \\ 
 &  to 10 16:30 & \textit{On November 7, Verizon stores released the new DROID phone, promoted as an iPhone alternative}\\
 
\hline
\multirow{ 2}{*}{23} & from 18 14:00 & \textbf{read}: blog (0.76), article (0.74) \\ 
 &  to 20 09:00 & \textit{No clear corresponding event}\\
 
\hline
\multirow{ 2}{*}{24} & from 02 05:00 & \textbf{wave}: guide (0.81), google (0.73) \\ 
 &  to 03 19:00 & \textit{The complete Google Wave guide was released on November 9}\\
 
\hline
\multirow{ 2}{*}{25} & from 18 10:30 & \textbf{talk, show}: oprah (0.89), 2011 (0.85), end (0.77) \\ 
 &  to 20 09:00 & \textit{Related to event \# 15}\\
 
\hline
\end{tabular} 
\end{footnotesize}
\label{tab:top25}
\end{table*}

Next, we qualitatively analyze the results of \textit{MABED} and show how they provide relevant information about the detected events. Table \ref{tab:top25} (page \pageref{tab:top25}) lists the top 25 events with highest magnitude of impact over the crowd in $\mathcal{C}_{en}$. From this table, we make several observations along three axes: readability, temporal precision and redundancy.

\noindent\textbf{Readability}~~ We argue that highlighting main words allows for an easy reading of the description, more especially as main words often correspond to named entities, \textit{e.g.} Fort Hood ($e_6$), Chrome ($e_7$), Tiger Woods ($e_8$), Obama ($e_{13}$). This favors a quicker understanding of events by putting into light the key places, products or actors at the heart of the events, in contrast with existing methods that identify bags of words or n-grams. What is more, \textit{MABED} ranks the words that describe each event and limits their number, which again favors the interpretation of events.

\pgfplotsset{height=0.6\columnwidth, width=1.00\columnwidth}
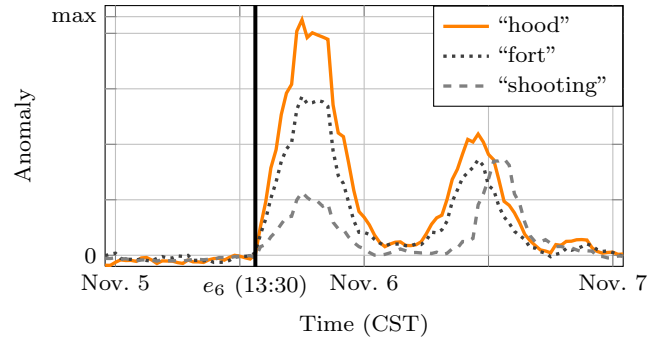
\begin{figure}[]
\centering
	\begin{tikzpicture}[]
		\begin{axis}[grid=major, ymin=-1, ymax=21.5, legend style={anchor=north east,legend cell align=left,at={(1,1)}},  ytick \empty, extra y ticks={0,5,10,15,20,21.5}, extra y tick labels={0, , , , ,max}, xtick \empty, extra x ticks={202,226,229,250,274,298}, extra x tick labels={Nov. 5, ,$e_6$ (13:30), Nov. 6, , Nov. 7, },
    xlabel={Time (CST)},
    ylabel={Anomaly},
    xmin=200,
    xmax=300,
    ymax=22.5
		]
			\addplot+[mark=empty, color=orange, line width=1.25pt] coordinates {
			(200,-0.8980885666965329)
(201,-0.9107868998314549)
(202,-0.6311050682639834)
(203,-0.4136450690576292)
(204,-0.4749155707044443)
(205,-0.4584077376290458)
(206,-0.5492008195437372)
(207,-0.24031282176594548)
(208,-0.202852739017926)
(209,-0.4272968214484872)
(210,-0.743803819107232)
(211,-0.5076148227976851)
(212,-0.519994653333417)
(213,-0.5282485698711161)
(214,-0.6625344870437321)
(215,-0.736184819226279)
(216,-0.4317412380457099)
(217,-0.595549735486202)
(218,-0.5961846521429481)
(219,-0.6457081513691436)
(220,-0.4130101524008831)
(221,-0.2945971516469196)
(222,-0.27427981863104456)
(223,-0.439676651984219)
(224,-0.18951948922625805)
(225,-0.002853992142906381)
(226,-0.0863465767758348)
(227,-0.2238070772321811)
(228,-0.3384105780456681)
(229,0.10508183950392957)
(230,2.6120659227281364)
(231,4.817462505456315)
(232,7.877144671190448)
(233,9.50889092173608)
(234,12.673652838432508)
(235,14.610161172757898)
(236,15.400636587490053)
(237,20.194605088105128)
(238,21.220001754374973)
(239,19.614922003412676)
(240,20.049213350793543)
(241,19.84825909612095)
(242,19.66667293229157)
(243,19.461592763620946)
(244,13.55333717624994)
(245,11.021907757470192)
(246,10.690160671507906)
(247,8.335240171845205)
(248,5.842224255069413)
(249,3.9434945060912314)
(250,3.425081923045595)
(251,2.738732840019799)
(252,1.8209576798808693)
(253,1.1593703464681715)
(254,0.7736553446824685)
(255,1.0092094243352694)
(256,1.17524117434519)
(257,1.2063520905257485)
(258,0.8876218402975753)
(259,0.8685743405951925)
(260,0.9460362612598501)
(261,1.2114335123213513)
(262,1.9371474280654046)
(263,2.792386430327295)
(264,3.3733393483332392)
(265,3.607625683214182)
(266,5.030800767747905)
(267,6.832704682301489)
(268,7.7561961808927675)
(269,9.138101682737993)
(270,10.44032409989077)
(271,10.210165767549494)
(272,10.921276600188387)
(273,10.206991184265764)
(274,9.087308350198306)
(275,8.6060373473015)
(276,6.801910180178487)
(277,4.858417762628889)
(278,4.500004177083269)
(279,3.8273047579066963)
(280,2.938733675436453)
(281,1.8898450928670036)
(282,1.3231788439086638)
(283,0.8907985121229399)
(284,0.48508259137891785)
(285,0.3806377570733684)
(286,0.22349484025789307)
(287,0.9631769224503606)
(288,0.8942915980058603)
(289,1.251114759097165)
(290,1.2669876755158174)
(291,1.3961942594344645)
(292,1.4139719258233554)
(293,1.373018757192415)
(294,0.6657174244939986)
(295,0.344447507638841)
(296,0.1879395074801118)
(297,0.25587558975194385)
(298,0.2831770059920259)
(299,0.08158992320432407)
(300,0.11524050601186708)
			};
			\addlegendentry{{\footnotesize ``hood''}}
			\addplot+[mark=empty, color=darkgray, line width=1.25pt, dotted] coordinates {
			(200,-0.04784013466916459)
(201,0.11030528213064672)
(202,0.1861444059456604)
(203,-0.27977059458691783)
(204,-0.3757108873586841)
(205,-0.364633262531772)
(206,-0.4255601990797886)
(207,-0.3301222005710073)
(208,-0.4716511801304503)
(209,-0.5104228670246427)
(210,-0.2776402821202039)
(211,-0.11914503459669218)
(212,-0.23929465771935426)
(213,-0.24483347013281037)
(214,-0.05643761630566224)
(215,-0.105860865533424)
(216,-0.013405304478042113)
(217,-0.4566627610938111)
(218,-0.45708882358715386)
(219,-0.49032169806789017)
(220,-0.44601119876024176)
(221,-0.6450586149209592)
(222,-0.6314246151339904)
(223,-0.6305724901473049)
(224,-0.4627038677702521)
(225,-0.17077482806080999)
(226,-0.11496064143290666)
(227,-0.09536176673913907)
(228,-0.06042464228503164)
(229,0.23498912914079484)
(230,1.906342483234232)
(231,3.2656656286614747)
(232,5.139048836369028)
(233,6.615615399235176)
(234,8.44767054508848)
(235,9.57173096242087)
(236,10.154813148622711)
(237,13.12839727403546)
(238,14.312106440435837)
(239,13.686767835623419)
(240,13.750130011716719)
(241,13.950805446081166)
(242,13.828951572985131)
(243,13.415016196640371)
(244,9.421756964764215)
(245,7.1791749425128915)
(246,6.458748169915394)
(247,4.777595378475072)
(248,3.115615399235176)
(249,2.0717309624208706)
(250,2.05937515011393)
(251,1.8795005461536372)
(252,1.4324847588674257)
(253,1.1025599254808345)
(254,0.8459178409492003)
(255,1.0039870259793693)
(256,1.0035609634860265)
(257,1.0244380256598224)
(258,0.8675707963400396)
(259,0.8547889215397565)
(260,1.0164190700559521)
(261,0.9708303832682753)
(262,1.3437875284824865)
(263,2.0799786133336675)
(264,2.022460176732393)
(265,2.2893277611042535)
(266,3.0711075327431114)
(267,4.828525510491789)
(268,5.8890263845464625)
(269,6.976445427451374)
(270,7.455568365277577)
(271,7.9129621159433)
(272,8.609453157143962)
(273,8.077498470143254)
(274,6.275693761317285)
(275,5.900104009373375)
(276,4.419702884067142)
(277,3.4576224459746494)
(278,2.8859677153234213)
(279,2.1604281928058096)
(280,1.7900274225516548)
(281,1.423887277230928)
(282,0.7124840487632702)
(283,0.6460182998017975)
(284,0.3211299845656773)
(285,0.19621744224660248)
(286,-0.1307249536865893)
(287,0.08493889971449634)
(288,0.20318690567937137)
(289,0.6093006936046766)
(290,0.45328558927157936)
(291,0.5948145688310224)
(292,0.7734109853112867)
(293,1.0244380256598224)
(294,0.8305033594192184)
(295,0.5052652134593978)
(296,0.3454155466862154)
(297,0.3910042334738921)
(298,0.2426582540209648)
(299,0.052558150220445576)
(300,-0.09152720429905412)
			};			
			\addlegendentry{{\footnotesize ``fort''}}
			\addplot+[mark=empty, color=gray, line width=1.25pt, dashed] coordinates {
			(200,-0.28310390703483485)
(201,-0.2904555735866317)
(202,-0.2250257412756395)
(203,-0.36228574173511874)
(204,-0.3012669093552046)
(205,-0.4583764095045353)
(206,-0.5109408253498829)
(207,-0.4286021599697579)
(208,-0.4069147436419572)
(209,-0.440364826452633)
(210,-0.5271144917638361)
(211,-0.3903734939004142)
(212,-0.4940319922807502)
(213,-0.4988105755394181)
(214,-0.4800638258323361)
(215,-0.522703491832758)
(216,-0.44293790974576186)
(217,-0.3711077415972749)
(218,-0.3714753249248648)
(219,-0.40014682447687255)
(220,-0.36191815840752883)
(221,-0.389854491304357)
(222,-0.21142515815481527)
(223,-0.21068999149963558)
(224,-0.23252882709590375)
(225,-0.1244593287844897)
(226,-0.07630591287022015)
(227,-0.05939707980108733)
(228,-0.029255246938720114)
(229,-0.22863682976088287)
(230,0.7754065868426054)
(231,0.9644958364922527)
(232,1.6657153359523644)
(233,2.145498252934923)
(234,2.81106216961882)
(235,3.4409705035265024)
(236,3.512649252406522)
(237,4.989859086095952)
(238,5.671229085866212)
(239,5.206884668642426)
(240,4.853053343441875)
(241,4.859518424070024)
(242,4.421056259045995)
(243,4.661974757885812)
(244,2.9782478388815443)
(245,2.389876420991514)
(246,2.320619336136156)
(247,1.6414548363314347)
(248,1.1454982529349231)
(249,0.7743038368598358)
(250,0.5969772536930638)
(251,0.25224883720444524)
(252,0.010379007650327954)
(253,0.013319674271046744)
(254,0.07948467323721864)
(255,0.2158580877730507)
(256,0.3821571711121275)
(257,0.5668354208306965)
(258,0.4086231706985962)
(259,0.2309290042042343)
(260,0.24945958985219385)
(261,0.21012817380008053)
(262,0.24431342326593602)
(263,0.16050442457545167)
(264,0.27754734201748943)
(265,0.05353767624680705)
(266,0.1757267602751027)
(267,0.5873553423850728)
(268,0.6395521749028307)
(269,1.2378483426377862)
(270,1.219836759585884)
(271,1.5164117601602327)
(272,4.542142593091523)
(273,5.5145738435222835)
(274,6.875108343097265)
(275,8.4824426747535)
(276,8.499351507822633)
(277,8.698733090644795)
(278,8.114037506030664)
(279,5.540369421244227)
(280,4.675056338410578)
(281,2.9800857555194935)
(282,1.774822839455893)
(283,1.2174798403518776)
(284,1.2018899213246366)
(285,1.4045797540951082)
(286,1.0767601706756222)
(287,0.785698920015121)
(288,0.7896775918279542)
(289,0.6629260886001132)
(290,0.6721156717898593)
(291,0.4837615887953919)
(292,0.32738725530124074)
(293,0.4001687541640298)
(294,0.3766434211982798)
(295,-0.11637249557751314)
(296,-0.11049116233607562)
(297,-0.23782641295062898)
(298,-0.2220203298642657)
(299,-0.07557074621504045)
(300,-0.056088829852778715)
			};
			\addlegendentry{{\footnotesize ``shooting''}}
			\addplot+[mark=empty, color= black, line width=1.5pt] coordinates {(229,50)(229,-10)};
		\end{axis}
	\end{tikzpicture}
	\caption{Measured anomaly for the words ``hood'', ``fort'' and ``shooting'' between Nov. 5 and Nov. 7 midnight (CST).}
	\label{fig:event6}
\end{figure}

\pgfplotsset{height=0.6\columnwidth, width=1.00\columnwidth}
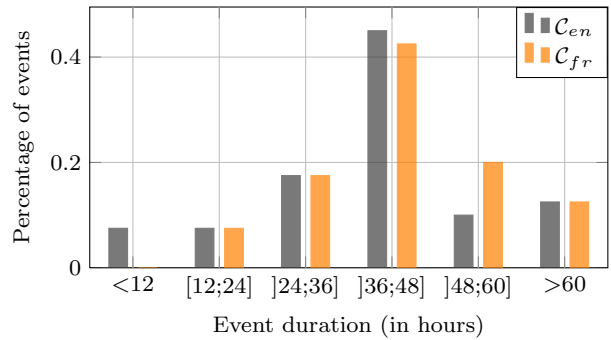
\begin{figure}[]
\centering
	\begin{tikzpicture}[]
		\begin{axis}[grid=major,ymin=0,  ybar, xtick \empty, extra x ticks={0,1,2,3,4,5}, extra x tick labels={$<$12, [12;24], ]24;36], ]36;48], ]48;60], $>$60}, bar width=7pt,legend style={anchor=north east, at={(1,1)}},
	ybar = 4pt,
    xtick align=inside,
    xlabel={Event duration (in hours)},
    ylabel={Percentage of events}]
    		\addplot+[color=darkgray, opacity=0.7] coordinates {
    		(0,0.075)
            (1,0.075)
            (2,0.175)
            (3,0.45)
            (4,0.1)
            (5,0.125)
            };
    		\addlegendentry{$\mathcal{C}_{en}$}
			\addplot+[color=orange, opacity=0.7] coordinates {
			(0,0)
            (1,0.075)
            (2,0.175)
            (3,0.425)
            (4,0.2)
            (5,0.125)
            };
			\addlegendentry{$\mathcal{C}_{fr}$}
		\end{axis}
	\end{tikzpicture}
	\caption{Distribution of the duration of the events detected with \textit{MABED}.}
	\label{fig:duration}
\end{figure}

\noindent\textbf{Temporal precision}~~ \textit{MABED} dynamically estimates the period of time during which each event is discussed on Twitter. This improves the temporal precision as compared to existing methods, that typically report events on a daily basis. We illustrate how this improves the quality of the results with the following example. The 6$^{\text{th}}$ event corresponds to Twitter users reporting the Fort Hood shooting that, according to Wikipedia\footnote{Source: \url{http://en.wikipedia.org/wiki/Fort_Hood_shooting}}, happened on November 5, 2009 between 13:34 and 13:44 CST (\textit{i.e.} 19:34 and 19:44 UTC). The burst of activity engendered by this event is first detected by \textit{MABED} in the time-slice covering the 19:30-20:00 UTC period. \textit{MABED} gives the following description: 

\noindent\textbf{(i)} \textit{11-05 19:30 to 11-08 9:00}; \textbf{(ii)} \textit{hood, fort}; \textbf{(iii)} \textit{ft (0.92), shooting (0.83), news (0.78), army (0.75), forthood (0.73)}.

\noindent We can clearly understand that (i) something happened around 7:30pm UTC, (ii) at the Hood Fort and that (iii) it is a shooting. In contrast, $\alpha$-\textit{MABED} fails at detecting this event on November 5 but reports it on November 7 when the media coverage was the highest.

\noindent\textbf{Redundancy}~~ Some events have several main words, \textit{e.g.} events $e_1$, $e_5$, $e_6$, $e_8$. This is due to merges operated by the third component of \textit{MABED} to avoid duplicated events. Redundancy is further limited because of the dynamic estimation of each event duration. We may continue using event $e_6$ to illustrate that. Figure \ref{fig:event6} plots the evolution of the anomaly measured for the words ``hood'', ``fort'' and ``shooting'' between November 5 and November 7. We see that the measured anomaly is closer to 0 during the night (local time), giving a ``dual-peak'' shape to the curves. Nevertheless, \textit{MABED} reports a unique event which is discussed for several days, instead of reporting distinct consecutive 1-day events. The importance of dynamically estimating the duration of events is further illustrated by Figure \ref{fig:duration}, which shows the distributions of event duration for both corpora. It reveals that some events are discussed during less than 12 hours whereas some are discussed for more than 60 hours. We note that event durations in $\mathcal{C}_{fr}$ are normally distributed and that these politics-related events tend to be discussed for a longer duration than the events detected in $\mathcal{C}_{en}$. This is consistent with the empirical study presented by \cite{romero_complex-contagion}, which states that controversial and more particularly political topics are more persistent than other topics on Twitter. 

\subsection{Analysis of Detected Events}

In this section we analyze events detected by \textit{MABED} in $\mathcal{C}_{en}$, with regard to the communities detected in the structure of the network that interconnects the authors of the tweets in that corpus. Our goal is to show that \textit{MABED} can help understanding users', or user communities' interests.

\noindent\textbf{Network structure}~~ The 52,494 authors of the tweets in $\mathcal{C}_{en}$ are interconnected by 5,793,961 following relationships \citep{kwak_twitter-graph}. This forms a directed, connected network, which diameter is 8, with an average path length of 2.55 and a clustering coefficient of 0.246. We measure a small-world-ness metric \citep{humphries_sigma-smallworld} of $s = 47.2$, which means -- according to the definition given by \cite{humphries_sigma-smallworld}, \textit{i.e.} $s > 1$ -- this network is a small-world network.

\noindent\textbf{Detecting communities in the network}~~ In order to find communities in this network, we apply the Louvain method proposed by \cite{blondel_louvain}, which has been notably used to detect communities in online social networks in several studies \citep{haynes_louvain-linkedin,kim_louvain-twitter}. It is a heuristic-based, greedy method for optimizing modularity \citep{newman_modularity}. It detects two communities: $c_0$, comprised of 25,625 users, and $c_1$, comprised of 26,869 users. By $\mathcal{C}_{en}(c_0)$ and $\mathcal{C}_{en}(c_1)$, we denote, respectively, the corpus of 479,899 tweets published by the users belonging to $c_0$ and the corpus of 932,699 tweets published by users belonging to $c_1$.

\noindent\textbf{Characterizing communities' interests}~~ We extract the list $L_0$ (respectively $L_1$) of the $k$ most impactful events from the corpus $\mathcal{C}_{en}(c_0)$ (respectively $\mathcal{C}_{en}(c_1)$). Here we choose to fix $k = 10$, so only events that engendered a significant amount of reactions in the related community are considered. In order to compare the two communities' interests, we first manually label each detected event with one the following categories \citep{mcminn_corpus}:
\begin{itemize}
\item Armed Conflicts and Attacks;
\item Sports;
\item Disasters and Accidents;
\item Art\, Culture and Entertainment;
\item Business and Economy; 
\item Law\, Politics and Scandals; 
\item Science and Technology;
\item Miscellaneous.
\end{itemize}

Then, we measure the weight of each category in each event list and compute the category weight distribution for each community. The weight of a given category is obtained by the following formula:
$$
\text{Weight}_{\text{category}} = \displaystyle \sum_{e \in E} 1 - ((\text{rank}(e) - 1) \times 0.1) 
$$
where $E$ is the set of events labelled with this category. Thus, the contribution of an event to the weight of the related category linearly diminishes with its rank in the list, \textit{e.g.} an event ranked 1$^{\text{st}}$ contributes for a weight of 1, whereas an event ranked 10$^{\text{th}}$ contributes for a weight of 0.1. 

\pgfplotsset{height=0.3\textwidth, width=1.0\columnwidth}
\begin{figure}[]
\begin{center}
	\begin{tikzpicture}[]
		\begin{axis}[grid=major,ymin=0,  ybar = 3pt, xtick \empty, 
		xtick align=inside,
		extra x ticks={7,6,5,4,3,2,1,0},
		extra x tick labels={}, 
	    bar width=5pt,
	    legend style={anchor=north east,legend cell align=left,at={(1,1)}},		
    ylabel={Weight}]
    		\addplot+[color=darkgray, opacity=0.7] coordinates {
    		(7,0)
            (6,0)
            (5,0.1)
            (4,0.4)
            (3,0.7)
            (2,0.8)
            (1,0.9)
            (0,2.6)
    		};
			\addlegendentry{$\mathcal{C}_{en}(c_0)$}
			\addplot+[color=orange, opacity=0.7] coordinates {
			(7,1.5)
            (6,0.7)
            (5,0.4)
            (4,0.5)
            (3,1)
            (2,0)
            (1,1.1)
            (0,0.3)
			};			
			\addlegendentry{$\mathcal{C}_{en}(c_1)$}
			\end{axis}
	\end{tikzpicture}
	
	(a) Events detected in $\mathcal{C}_{en}(c_0)$ and $\mathcal{C}_{en}(c_1)$.

    \vspace{0.3cm}
	
	\begin{tikzpicture}[]
		\begin{axis}[grid=major,ymin=0,  ybar = 3pt, xtick \empty, 
		extra x ticks={7,6,5,4,3,2,1,0}, 
		xtick align=inside,
        extra x tick labels={
        Miscellaneous, 
        Armed Conflicts and Attacks,
        Sports,
        Disasters and Accidents,
    	Art\, Culture and Entertainment,
	    Business and Economy, 
	    Law\, Politics and Scandals, 
	    Science and Technology},	
	    bar width=5pt,
	    x tick label style={rotate=90,anchor=east},
	    legend style={anchor=north east,legend cell align=left,at={(1,1)}},	
        ylabel near ticks,
        xlabel near ticks,
    xlabel={Event category},
    ylabel={Weight}
    ]
    		\addplot+[color=darkgray, opacity=0.7] coordinates {
    		(7,0.8)
            (6,0.5)
            (5,0)
            (4,0.3)
            (3,2.7)
            (2,0)
            (1,0.6)
            (0,0.6)
    		};
			\addlegendentry{$\mathcal{C}_{en}$}
			\addplot+[color=orange, opacity=0.7] coordinates {
			(7,1.2)
            (6,0.6)
            (5,0)
            (4,0.5)
            (3,1.9)
            (2,0)
            (1,0.8)
            (0,0.5)
            };
			
			\addlegendentry{$\mathcal{C}_{en}(\text{random})$}
			\end{axis}
	\end{tikzpicture}
	
    (b) Events detected in $\mathcal{C}_{en}$ and $\mathcal{C}_{en}(\text{random})$.
    
\end{center}
	\caption{Category weight distribution for events detected with \textit{MABED} in $\mathcal{C}_{en}(c_0)$, $\mathcal{C}_{en}(c_1)$, $\mathcal{C}_{en}$ and $\mathcal{C}_{en}(\text{random})$.}
	\label{fig:category_distribution}
\end{figure}
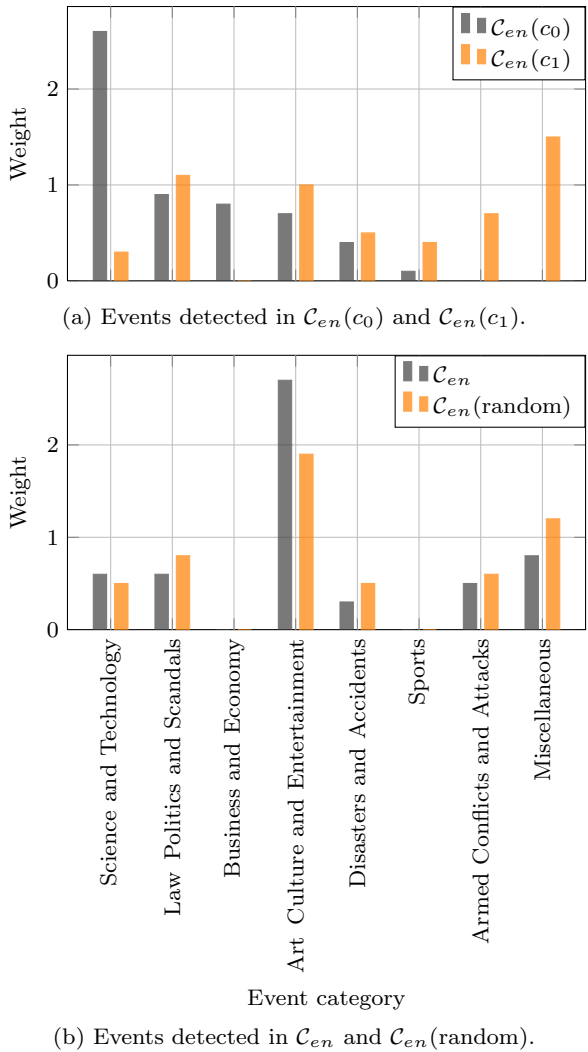

For comparison purpose, we do the same for events detected in $\mathcal{C}_{en}$ and $\mathcal{C}_{en}(\text{random})$. The $\mathcal{C}_{en}(\text{random})$ corpus contains 725,806 tweets that corresponds to the subset of tweets published by 26,000 randomly chosen authors (\textit{i.e.} about the number of users in communities $c_0$ or $c_1$) in $\mathcal{C}_{en}$. Figure \ref{fig:category_distribution}(a) shows the weight distribution for communities $c_0$ and $c_1$. We note that, visually, the two distributions seem quite different. For instance, we notice that the ``Miscellaneous'' category has the highest weight for $c_1$ whereas none of the events detected in $\mathcal{C}_{en}(c_0)$ belongs to that category. On the opposite, the ``Science and Technology'' category has the highest weight for $c_0$ whereas it has the second lowest weight for $c_1$. We measure a negative linear correlation of -0.36 between the two distributions using Pearson's coefficient, which reinforce this assessment. Figure \ref{fig:category_distribution}(b) shows the distribution of event category weight for $\mathcal{C}_{en}$ and $\mathcal{C}_{en}(\text{random})$. In this case, we measure a linear correlation of 0.93, which means that the two distributions are very similar. These results indicate that the communities detected in the network structure are also relevant in terms of users' interests. These results also shed light on the interplay between the social structure, \textit{i.e.} who follows whom, and the topical structure, \textit{i.e.} who's interested in what, in Twitter. More specifically, they complement the findings from \cite{romero_interplay-social-topical} -- who have found a relationship between the hashtags users adopt and their social ties -- and suggest that the social network structure can influence event detection. From a different perspective, these results show that event detection can help understanding user communities' interests.

\section{Implementation and Visualizations}

We provide a parallel implementation\footnote{Binaries: \url{http://mediamining.univ-lyon2.fr/mabed}}\footnote{Sources: \url{https://github.com/AdrienGuille/MABED}} of \textit{MABED}. It is also included in \textit{SONDY} \citep{guille_sondy}, an open-source social media data mining software that implements several state-of-the-art methods for event detection in social media. To ensure an efficient exploration of the events detected by \textit{MABED}, we also develop three visualizations, which we describe below.

\begin{figure*}[]
\begin{center}
\includegraphics[width=\textwidth]{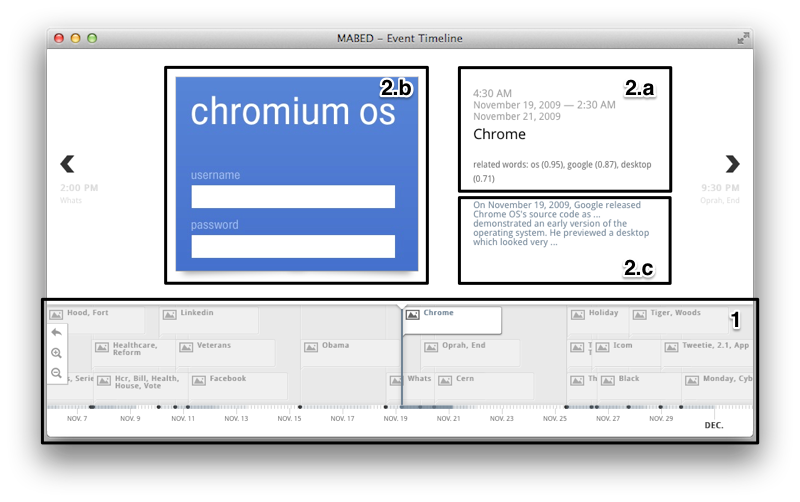}
\caption{The time-oriented visualization. The lower part is a ribbon (1) labeled with events in chronological order. The upper part displays details about the event selected in the lower part: the description extracted with \textit{MABED} (2.a), an image (2.b) and a hypertext (2.c).}
\label{fig:viz-timeline}
\end{center}
\end{figure*}

\noindent\textbf{Time-oriented visualization}~~ It is based on an interactive timeline that allows the user to explore the detected events through time. As an example, Figure \ref{fig:viz-timeline} shows the time-oriented visualization generated from the events detected in $\mathcal{C}_{en}$. As one can see, the timeline is divided into two parts. The lower part is a ribbon labelled with events in chronological order (fig. \ref{fig:viz-timeline}.1). Selecting a label (\textit{i.e.} the main term of an event) in the lower part updates the upper part of the visualization with details about the related event. More specifically, the upper part displays the temporal and textual descriptions extracted with \textit{MABED} (fig. \ref{fig:viz-timeline}.2.a), an image (fig. \ref{fig:viz-timeline}.2.b) and a hypertext (fig. \ref{fig:viz-timeline}.2.c). In the current implementation, these correspond to the top image and the description of the top page returned by the Bing search engine, using the description extracted by \textit{MABED} as a query. 

This visualization provides a chronological overview of the events detected in a tweet corpus. In addition, the hypertexts offer quick access to resources that can help learning more about these events. For instance, the hypertext associated to the event selected on the timeline depicted in Figure \ref{fig:viz-timeline}, reveals that, on November 19, 2009, Google released Chrome OS's source code and demonstrated an early version of this operating system for desktop computers.

\begin{figure*}[]
\begin{center}
\includegraphics[width=\textwidth]{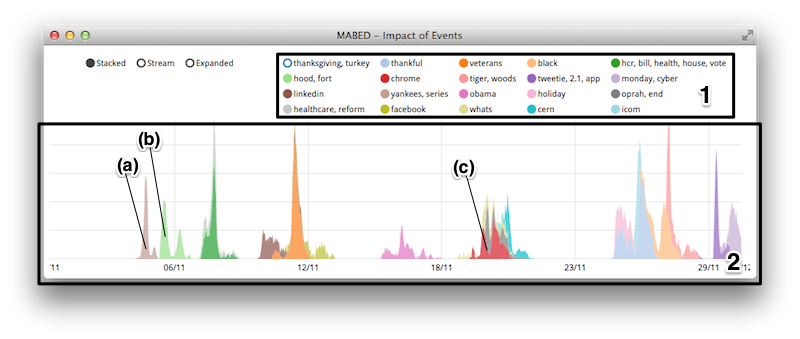}
\caption{The impact-oriented visualization is a chart (1) that plots the magnitude of impact of the detected events (2). Each event is associated to a different color.}
\label{fig:viz-impact}
\end{center}
\end{figure*}

\noindent\textbf{Impact-oriented visualization}~~ It is an interactive chart that allows analyzing the magnitude of impact of the detected events. More precisely, it plots the mention-anomaly function related to each event. Figure \ref{fig:viz-impact} depicts this visualization for the top 20 events detected in $\mathcal{C}_{en}$. The interactive legend (fig.\ref{fig:viz-impact}.1) can be used to display (fig.\ref{fig:viz-impact}.2) only the functions which are of interest by clicking on them. Single clicking on a legend item removes the corresponding function from the chart while double clicking on a item makes it the only visible one in the chart.

This visualization helps analyzing the temporal patterns that describe how Twitter users reacted to the detected events. For instance, we observe that different events trigger different patterns: some events engender a single significant peak of reactions (\textit{e.g.} fig.\ref{fig:viz-impact}.a), some events generate successive peaks of decreasing strength (\textit{e.g.} fig.\ref{fig:viz-impact}.b), while other events engender successive increasing peaks of attention (\textit{e.g.} fig.\ref{fig:viz-impact}.c).

\begin{figure*}[]
\begin{center}
\includegraphics[width=0.725\textwidth]{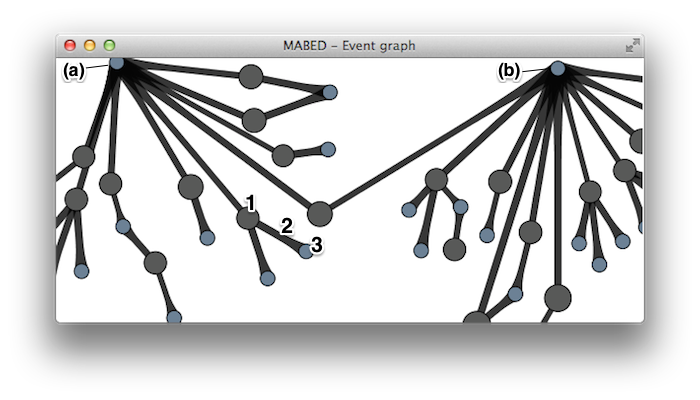}
\caption{The topic-oriented visualization is an interactive drawing of the event graph. Grey nodes (1) correspond to main terms while blue nodes (3) correspond to related words. Edges (2) connect words that describe the same event.}
\label{fig:viz-eventgraph}
\end{center}
\end{figure*}

\noindent\textbf{Topic-oriented visualization}~~ It is based on the event graph constructed by \textit{MABED} during the second phase. Figure \ref{fig:viz-eventgraph} shows the event graph constructed from $\mathcal{C}_{fr}$. Main terms are represented with grey nodes (fig. \ref{fig:viz-eventgraph}.1), whose diameter is proportional to the magnitude of impact of the corresponding event. Related words are represented by blue nodes (fig. \ref{fig:viz-eventgraph}.3), which are connected to main terms by edges (fig. \ref{fig:viz-eventgraph}.2) whose thickness is proportional to the related weight. For the sake of readability, nodes' labels are hidden by default, but the user can click a grey node in order to reveal the main term and related words describing the event.

This visualization helps identifying similar events by topic. It also helps discovering words which are common to several events. This could be useful in cases when one wants to quickly identify events involving, \textit{e.g.} a specific actor or place. For instance, we spot two nodes (fig.\ref{fig:viz-eventgraph}.a, b) which describe many events. They correspond to the two main candidates for the 2012 presidential elections. Interestingly, even though they appear in many different events, they appear together in only one single event.

\begin{figure*}[]
\begin{center}
\includegraphics[width=\textwidth]{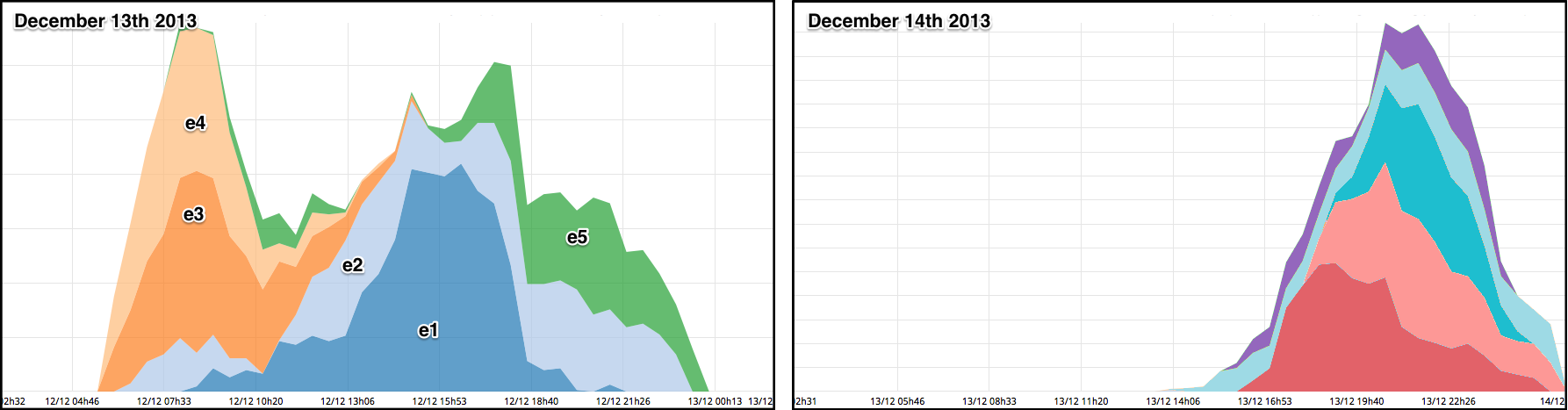}
\caption{On the left: the impact-oriented visualization generated based on the five most impactful events detected from tweets collected on December 13$^{\text{th}}$, 2013. On the right: the impact-oriented visualization generated based on the five most impactful events detected from tweets collected on December 14$^{\text{th}}$, 2013. Each event is associated to a different color.}
\label{fig:online-use-case}
\end{center}
\end{figure*}

\subsection{Case study: Monitoring the French Political Conversation on Twitter}
\label{sec:case-study}

\noindent\textbf{Setup}~~ \textit{MABED} has been used from December 2013 until November 2014 to continuously analyze the French political conversation on Twitter. For this purpose, it has been coupled with a system fetching tweets in real-time about the President of the French Republic, Francois Hollande, via the Twitter streaming API. During this period, the highest crawling rate we have reached was about 150,000 tweets in 24 hours. Every 10 minutes, the visualizations were refreshed with the five most impactful events detected from the tweets received in the last 24 hours (partioned into 144 time-slices of 10 minutes each).

\noindent\textbf{Results}~~ It has helped us to understand the political conversation on Twitter, and it has revealed interesting clues about how public opinion evolves on Twitter. The visualizations, most notably the impact-oriented visualizations in this case, have also shed light on specific patterns. As an example, Figure \ref{fig:online-use-case} shows two impact-oriented visualizations. They were automatically generated from the five most impactful events detected in two successive 24-hour periods: (i) December 13$^{\text{th}}$, 2013 and (ii) December 14$^{\text{th}}$, 2013. We notice that the two distinct sets of detected events are distributed differently throughout each day. On December 13$^{\text{th}}$, we observe that Twitter users focus their attention on various successive events throughout the day. On the other hand, on the 14$^\text{th}$ we observe an opposite pattern since all the events are discussed simultaneously during the second half of the day. Further analysis, indicates that all five events detected on December 13$^{\text{th}}$ are closely related. In contrast, the events detected on the 14$^{\text{th}}$ are loosely related. One possible explanation for the pattern we observe on December 13$^{\text{th}}$ is that related events can ``compete'' or ``collaborate''. The mention-anomaly measured for competing events should thus be negatively correlated, while the mention-anomaly measured for cooperating events should be positively correlated. For instance, we observe that events $e3$ and $e4$ take place simultaneously and are likely to collaborate. Eventually, both of them end when events $e2$ and $e1$ (likely to be competing with $e3$ and $e4$) start.

\section{Conclusion}
\label{sec:conclusion}

In this paper, we developed \textit{MABED}, a mention\hyp{}anomaly\hyp{}based method for event detection in Twitter. In contrast with prior work, \textit{MABED} takes the social aspect of tweets into account by leveraging the creation frequency of mentions that users insert in tweets to engage discussion. Our approach also differ from prior work in that it dynamically estimates the period of time during which each event is discussed on Twitter. The experiments we conducted have shown that \textit{MABED} has a linear runtime in the corpus size. They have also demonstrated the relevance of our approach. Quantitatively speaking, \textit{MABED} yielded better performance in all our tests than $\alpha$-\textit{MABED} -- a variant that ignores mentions -- and also outperformed two recent methods from the literature. Qualitatively speaking, we have shown that the highlighting of main words improves the readability of the descriptions of events. We have also shown that the temporal information provided by \textit{MABED} is very helpful. On the one hand, it clearly indicates when real-world events happened. On the other hand, dynamically identifying the period of time during which each event is discussed limits the fragmentation of events. By analyzing the detected events with regard to the user communities detected in the social network structure, we have shed light on the interplay between social and topical structure in Twitter. In particular, we have found that \textit{MABED} can help understanding user communities' interests. Moreover, we presented three visualizations designed to help with the exploration of the detected events. Finally, we described how we leveraged \textit{MABED} and the visualizations we developed in order to continuously monitor and analyze the French political conversation on Twitter from December 2013 until November 2014.

As part of future work, we plan to investigate the effectiveness of utilizing more features to model the discussions between users (\textit{e.g.} number of distinct users, users' geolocations). Another interesting direction for future work is to incorporate sentiment analysis in the event detection process to further enrich event descriptions.

\begin{acknowledgements}
The authors would like to thank Michael Gauthier for helpful suggestions.
This work was supported in part by the French National Research Agency and the ImagiWeb project (ANR-2012-CORD-002-01).
\end{acknowledgements}

\bibliographystyle{spbasic}      
\bibliography{snam-final}   

\end{document}